\pdfoutput=1
\documentclass[12pt]{article}
\usepackage{amsmath, amsthm, amsfonts, graphicx, lineno, caption}
\newcommand\apj{Astrophys. J.}
\newcommand\apjl{Astrophys. J. Lett.}
\newcommand\solphys{Sol. Phys.}
\newcommand\nat{Nature}
\newcommand\ssr{Space Sci. Rev.}
\newcommand\jgr{J. Geophys. Res.}
\newcommand\aap{Astron. \& Astrophys.}
\newcommand\aapr{Astron. \& Astrophys. Rev.}

\renewcommand{\vec}[1]{ {\mathbf #1} }

\newcommand{\grad}{ {\bf \nabla } }

\newcommand{\Fig}{{Fig.}}

\newcommand{\EFig}{Extended Data Fig.}
\newcommand{\SFig}{Supplementary Fig.}
\newcommand{\Movie}{Supplementary Video}

\newcommand{\dive}{\nabla\cdot}

\newcommand{\JB}{J\Delta/B}

\newcommand{\citep}{\cite}

\begin{document}

\title{\textbf{A Fundamental Mechanism of Solar Eruption Initiation}}

\author
{Chaowei Jiang,$^{1\ast}$ Xueshang Feng,$^{1\ast}$ Rui Liu,$^{2,3}$ XiaoLi Yan,$^4$\\
Qiang Hu,$^{5,6}$ Ronald L. Moore,$^5$ Aiying Duan,$^7$ \\
Jun Cui,$^7$ Pingbing Zuo,$^1$ Yi Wang,$^1$ Fengsi Wei$^1$ \\
\\
\small{$^{1}$Institute of Space Science and Applied Technology, }\\
\small{Harbin Institute of Technology, Shenzhen 518055, China}\\
\small{$^{2}$CAS Key Laboratory of Geospace Environment, }\\
\small{Department of Geophysics and Planetary Sciences, }\\
\small{University of Science and Technology of China, Hefei 230026, China}\\
\small{$^{3}$CAS Center for Excellence in Comparative Planetology, Hefei 230026, China}\\
\small{$^{4}$Yunnan Observatories, Chinese Academy of Sciences, Kunming 650216, China}\\
\small{$^{5}$Center for Space Plasma and Aeronomic Research,}\\
\small{The University of Alabama in Huntsville, Huntsville, AL 35899, USA}\\
\small{$^{6}$Department of Space Science, The University of Alabama in Huntsville, Huntsville, AL 35899, USA}\\
\small{$^{7}$School of Atmospheric Sciences, Sun Yat-sen University, Zhuhai 519000, China}\\
\\
\small{$^\ast$To whom correspondence should be addressed;}\\
\small{ E-mail:  chaowei@hit.edu.cn, fengx@spaceweather.ac.cn.}
}

\maketitle

\abstract{
  \textbf{Solar eruptions are spectacular magnetic explosions in the
    Sun's corona and how they are initiated remains
    unclear. Prevailing theories often rely on special magnetic
    topologies which, however, may not generally exist in the
    pre-eruption source region of corona. Here using fully
    three-dimensional magnetohydrodynamic simulations with high
    accuracy, we show that solar eruption can be initiated in a single
    bipolar configuration with no additional special topology. Through
    photospheric shearing motion alone, an electric current sheet
    forms in the highly sheared core field of the magnetic arcade
    during its quasi-static evolution. Once magnetic reconnection sets
    in, the whole arcade is expelled impulsively, forming a
    fast-expanding twisted flux rope with a highly turbulent
    reconnecting region underneath. The simplicity and efficacy of
    this scenario argue strongly for its fundamental importance in the
    initiation of solar eruptions.}
}

%\keywords{Sun: Coronal mass ejections; Magnetohydrodynamics (MHD);Methods: numerical;Sun: corona;Sun: flares}

\section{Introduction}
\label{sec:intro}

From time to time, the Sun produces eruptive activities, such as solar
flares and coronal mass ejections (CMEs). It is now confirmed that
such eruptions are explosive releases of magnetic energy in the Sun's
corona \cite{Fleishman2020}. Research on solar eruptions has a long
history of more than a century, from which a basic physical picture
has been established \cite{Priest2002, Forbes2006}. Prior to
eruptions, the coronal magnetic field is line-tied at the solar
surface (i.e., the photosphere), and is continuously but rather slowly
stressed by motions at the photosphere (such as surface shear and
rotational flows) that could last for a few hours or even days, during
which magnetic free energy accumulates. Since plasma is
strongly-magnetized in the corona, the Lorentz force dominates and is
mostly self-balanced, that is, the outward magnetic pressure of the
low-lying, strongly stressed flux is cancelled by the inward magnetic
tension of the overlying, mostly un-sheared flux. At a critical point,
there is a catastrophic disruption of this force balance, and the free
magnetic energy is rapidly converted into impulsive heating and fast
acceleration of the plasma. If the overlying flux is not too strong,
the eruptive magnetic field can successfully eject into the
heliosphere, forming a CME. Otherwise, it fails to escape, resulting
in a confined flare or failed eruption.

A fundamental question that still lies in this picture, also a central
point of controversy, is how the pre-eruption force balance is
abruptly destroyed. Due to the lack of regular measurements of
magnetic field in the corona, the mechanism governing the initiation
of solar eruptions has remained a subject of intense investigation for
decades
\cite{Forbes2006,Shibata2011,ChenP2011,Schmieder2013,Aulanier2014,Janvier2015}. The
existing theories fall into two categories; one is based on ideal
magnetohydrodynamic (MHD) instability and the other on a resistive
process, i.e., magnetic reconnection. The first category requires the
preexistence of a magnetic flux rope (MFR), a group of twisted
magnetic field lines winding tightly enough about a common axis, in
the corona before eruption, for the triggering of eruption to be a
loss of equilibrium \cite{LinJ2015} or ideal instabilities of the MFR,
such as torus instability
\cite{Kliem2006,Torok2005,Fan2007,Aulanier2010,Amari2018}. On the
other hand, the most prevailing model based on magnetic reconnection,
namely, the breakout model
\cite{Antiochos1999,Aulanier2000,Lynch2008,Wyper2017}, relies on a
multipolar magnetic configuration in which there must be a magnetic
null point above the sheared magnetic flux, such that reconnection at
the null can remove the overlying restraining flux to trigger an
eruption.

However, the key prerequisite of magnetic topology for these models,
either an MFR or a null point, may not generally exist in the source
region of eruptions, in particular, in solar active regions (ARs). For
the MFR-based models, although there is little doubt that MFR
constitutes the core structure of CMEs, the existence of MFR before
CME initiation is still in intense debates
\cite{Patsourakos2020}. Almost all the observed features that have
been invoked to support the preexistence of MFRs, such as coronal
sigmoids and filaments, can also fit in simply sheared arcades with
only weak twist \cite{DeVore2000,ChenP2011}. Furthermore, significant
magnetic twist is rarely seen before but only observed during filament
eruptions \cite{WangH2015,WangW2017}. For the breakout model, the
coronal null point, which exists primarily in multipolar magnetic
configuration, is not universally present in ARs, considering that the
commonly-seen ARs are a bipolar configuration consisting of a pair of
sunspots with opposite polarities. Furthermore, in this large,
multipolar magnetic field, the null point must be situated right above
the sheared arcade concentrated around the polarity inversion line
(PIL) such that the breakout reconnection can be effective enough,
which is difficult to fulfill in reality \cite{Ugarteurra2007}.

Here, with an ultra high-resolution, fully three-dimensional (3D) MHD
simulation, we show that solar eruptions can be initiated from a more
universal bipolar magnetic configuration without the aforementioned
special topology. The simulation covers the whole process from the
energy accumulation in the source region to the triggering of eruption
and its subsequent evolution. It shows that with surface shear along
the PIL of a bipolar field, a vertical current sheet (CS) can
spontaneously form above the PIL, essentially between the
strongly-sheared legs of the core of the magnetic arcade. Once the CS
is sufficiently thin such that ideal MHD is broken down, reconnection
sets in and triggers the eruption. The magnetic topology does not
change before the eruption, but transforms to a complex one having a
highly twisted erupting MFR after the eruption onset.
Although such scenario appears to be similar to an early proposed idea, namely the runaway tether-cutting reconnection model \cite{Moore1980,Moore1992,Moore2001} which was originally surmised from observation, that model has never been accomplished in 3D simulations and thus remains a conjectural ``cartoon''. Moreover, our simulation shows that the reconnection not only cuts the magnetic tethers, but also results in strong upward tension force, and it is the latter that plays the key role in driving the eruption.

\section{Results}
\label{sec:res}

\begin{figure*}[htbp]
  \centering
  \includegraphics[width=\textwidth]{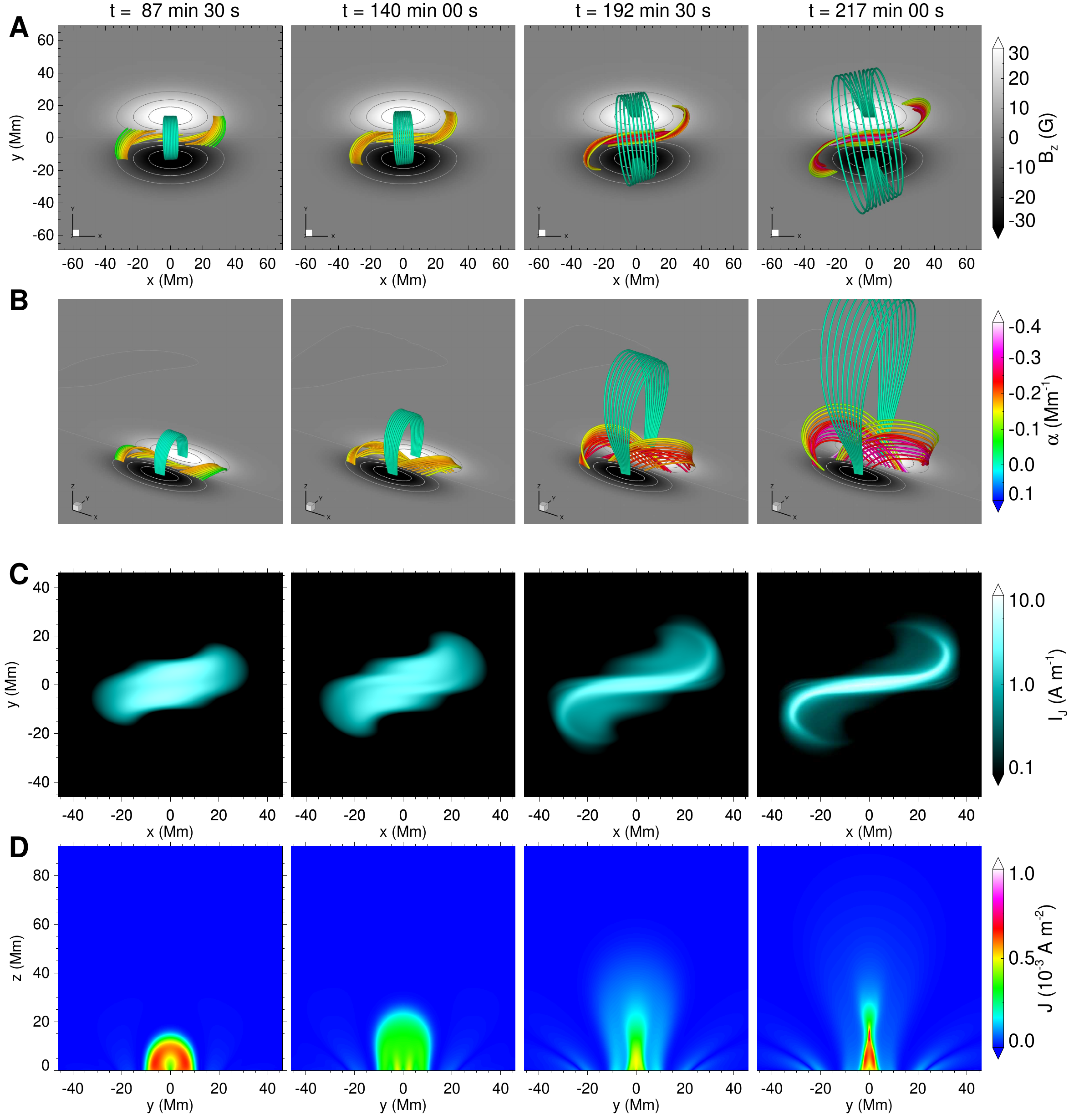}
  \caption{\textbf{Evolution of magnetic field lines and electric
      currents prior to eruption}. \textbf{(A)} Top view of magnetic
    field lines. The colored thick lines represent magnetic field
    lines and the colors denote the value of nonlinear force-free
    factor defined as $\alpha = \mathbf{J}\cdot \mathbf{B}/B^2$, which
    indicates how much the field lines are non-potential. In
    particular, for a perfectly force-free field, this parameter is
    constant along any given field line. As can be seen, the magnetic
    field is close to force-free since the color is nearly the same
    along any single field line. Note that at all different times, the
    field lines are traced from the same set of seeds at the bottom
    surface convected with the surface flow. The background shows the
    magnetic flux distribution on the bottom boundary (i.e., plane of
    $z=0$), and contours of $B_{z}=(-30,-20,-10,0,10,20,30)$~G are
    shown. \textbf{(B)} 3D prospective view of the same field lines
    shown in panel \textbf{(A)}. \textbf{(C)} Vertical integration of
    current density, i.e., $I_{J} = \int J dz$. \textbf{(D)} Vertical
    cross section (i.e., the $x=0$ slice) of current density.}
  \label{F1}
\end{figure*}

% pre-eruption
Our simulation solves the full MHD equations with both coronal plasma
pressure and solar gravity included. It is started with a potential,
bipolar magnetic field that mimics a typical solar AR consisting of a
pair of sunspots with opposite magnetic polarities (Fig.~\ref{F1}). In
particular, the flux distribution has a relatively strong-gradient and
elongated PIL, which is a characteristic magnetic field pattern
for eruption-productive ARs~\cite{Schrijver2007, Toriumi2019}.
%strong gradientacross the PIL, which is often observed in delta sunspots \cite{Toriumi2019} or colliding magnetic polarities \cite{Chintzoglou2019}.
The initial background atmosphere is set in
hydrostatic state and configured to simulate typical coronal
environment with low plasma $\beta$ (i.e., ratio of gas pressure to
magnetic pressure) and high Aflv{\'e}n speed (\EFig~1). Then we
energize the MHD system by applying anti-clockwise rotational flow to
both polarities at the bottom surface. Such a flow follows the
contours of the magnetic flux and concentrates near the PIL, thus it
preserves the magnetic flux distribution and produces strong shear
along the PIL. The flow speed, a few kilometers per second, is smaller
than the sound speed by two orders and the Alfv{\'e}n speed by three
orders of magnitude, respectively, thus representing a quasi-static
stress of the coronal magnetic field.

Figure~\ref{F1} and \Movie~1 show evolution of magnetic field lines
and electric current structure driven by the rotational flow. Since the
surface velocity has the largest gradient across the PIL, it creates
strongly-sheared, low-lying magnetic field lines there, and the
overall structure resembles an inverse S shape. In details, these
highly sheared field lines consist of two groups of J shape having an
oppositely curved elbow on their ends, and their arms are sheared past
each other above the middle of the PIL, where the electric current is
the strongest. Initially the current is volumetric but later it is
squeezed into a vertical, narrow layer extending above the PIL as an
inverse S shape (Fig.~\ref{F1}C and D), which is reminiscent of hot
sigmoid structures often observed prior to flare in the corona. On the
other hand, field lines connecting the central parts of the magnetic
polarities (analogous to sunspot's umbra) are only weakly sheared,
which play the role of the strapping field overlying the inner
strongly-sheared core. Clearly, the whole magnetic configuration
inflates during the energizing phase as magnetic pressure of the
sheared arcade increases continuously, which then stretches outward
its overlying field, making the bipolar arcade tend to approach an
open field configuration. But the magnetic field is still close to
force-free, i.e., the outward magnetic pressure gradient is balanced
by the inward magnetic tension force, and furthermore, the strapping
field is close to current-free, although it has been greatly
strengthened compared with the initial potential field
(\EFig~2).

\begin{figure*}[htbp]
  \centering
  \includegraphics[width=0.6\textwidth]{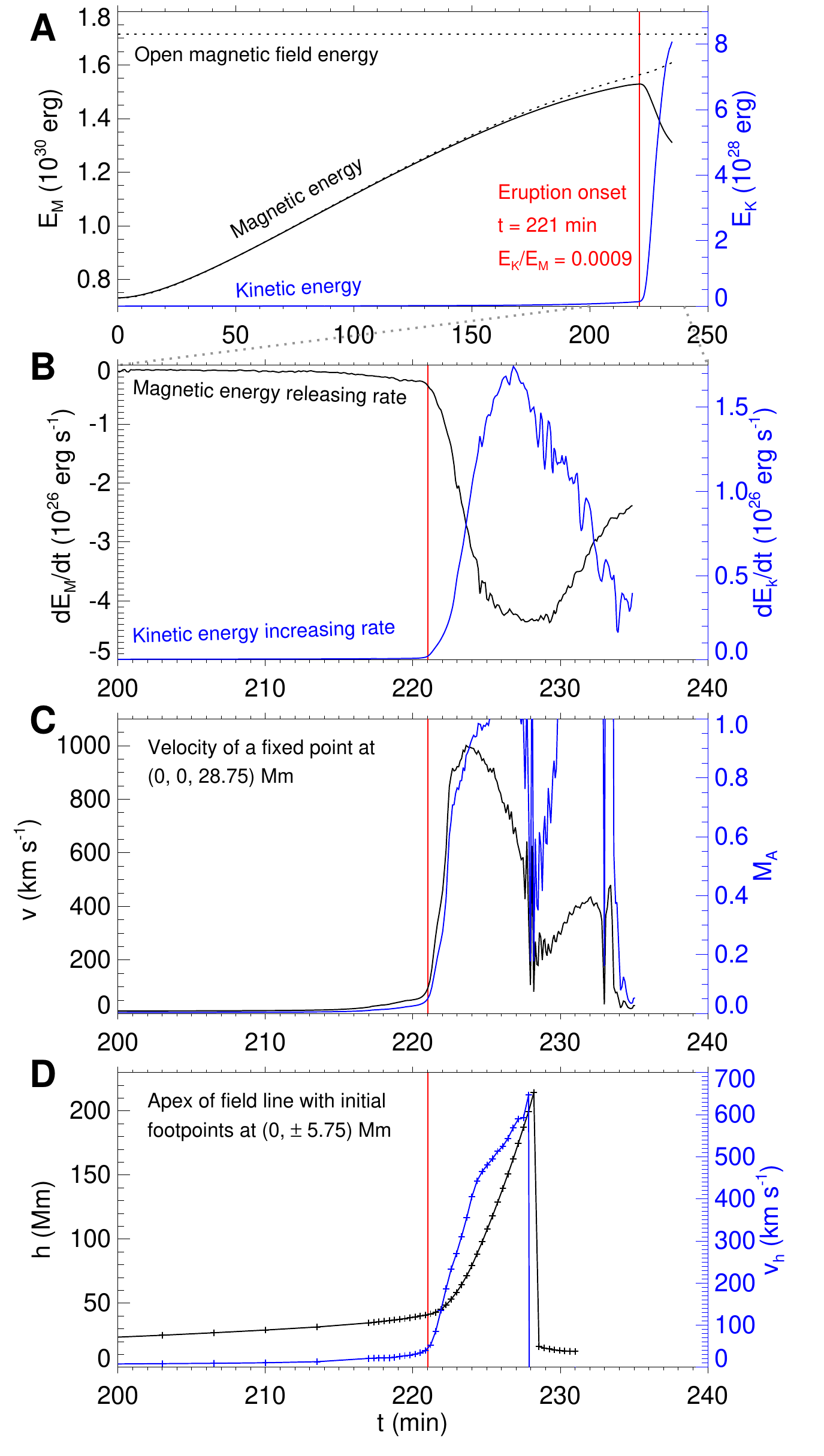}
  \caption{\textbf{Temporal evolution of different parameters in the
      simulations.}  \textbf{(A)} Evolution of magnetic energy $E_{M}$
      (the black line)
    and kinetic energy $E_{K}$ (the blue line). The dashed curve shows the energy
    injected into the volume (i.e., time integration of total Poynting
    flux) from the bottom boundary through the surface flow. The
    horizontal dashed line denotes the value of the total magnetic
    energy for a open force-free field with the same magnetic flux
    distribution on the bottom surface. \textbf{(B)} Releasing rate of
    magnetic energy (the black line) and increasing rate of kinetic
    energy (the blue line). \textbf{(C)} Magnitude of velocity (the black line) and Aflv{\'e}nic Mach
    number (the blue line) at a fixed point with location of $(x,y,z)=(0,0,28.75)$~Mm.
    \textbf{(D)} Height (the black line) and rising speed (the blue line) of the apex of a magnetic
    field line in the AR's core initially overlying the current
    sheet. In all panels, the red lines are shown for denoting the
    transition time (at $t=221$~min) from pre-eruption to eruption.}
  \label{F2}
\end{figure*}

The surface flow continuously injects magnetic energy into the
simulation volume (Fig.~\ref{F2}A). In the early phase, the magnetic
energy increases almost linearly, while the kinetic energy is
negligible. Thus almost all the energy brought by the surface flow
through Poynting flux is stored in the magnetic field. From around
$t=150$~min to $220$~min, the kinetic energy shows a slow rise
(Fig.~\ref{F2}A and B), indicating that a small amount of magnetic
energy is converted into kinetic energy and gravitational potential
energy of the plasma as it expands with the magnetic field. But the
system is still quasi static as the kinetic energy remains to be three
orders of magnitude less than the magnetic energy, and the velocity in
the core of the AR is less than the local Alfv{\'e}n speed by two
orders of magnitude (Fig.~\ref{F2}C).

% eruption
A sharp transition of the evolution pattern, namely an eruption,
occurs at $t=221$~min as the kinetic energy increases impulsively by
nearly two orders of magnitude in about 10~min, amounting to
$\sim 5$\% of the original magnetic energy. Meanwhile, the magnetic
energy drops immediately, despite the continual injection of Poynting
flux through the bottom surface, indicating that the magnetic energy
releases quickly during the eruption. The transition time, i.e., the
onset time of eruption, is more distinctly shown by time profiles of
the magnetic energy releasing rate (which is the substraction of the
magnetic energy changing rate from the total Poynting flux) and the
kinetic energy increasing rate. Both of them increase sharply at the
eruption onset, and reach their peaks simultaneously at around
$t=227$~min. The kinetic energy gained by plasma (mainly the CME) accounts
for approximately one third of the amount of released magnetic energy, indicating
that the flare energy should consume the other two thirds.
This is consistent with the energy partition between flare and CME in
typical eruptive flares \cite{Emslie2012}.
Also such a synchronization
of the evolutions of flare energy releasing rate and CME acceleration agrees well with observations
\cite{ZhangJ2001}. The impulsiveness of the
eruption is further seen from the evolution of velocity at a fixed
point in the AR core (Fig.~\ref{F2}C), which increases by more than 10
times in 2~min, reaching an Alfv{\'e}nic speed of
$\sim 1000$~km~s$^{-1}$. We further trace the rising apex of a single
field line (corresponding to a coronal loop in observation) initially
in the AR core (Fig.~\ref{F2}D). Prior to the eruption, the loop rises slowly
with velocity of $\sim 10$~km~s$^{-1}$, and once the eruption is
triggered, it ascends exponentially, gaining a speed of
$\sim 700$~km~s$^{-1}$ in about 5~min before its reconnection with
others during the eruption. Such a transition from slow-rise to
fast-acceleration phases is frequently observed for ejecting hot
coronal loops and filaments in eruptions \cite{ZhangJ2006,ChengX2020}.

\begin{figure*}[htbp]
  \centering
  \includegraphics[width=1\textwidth]{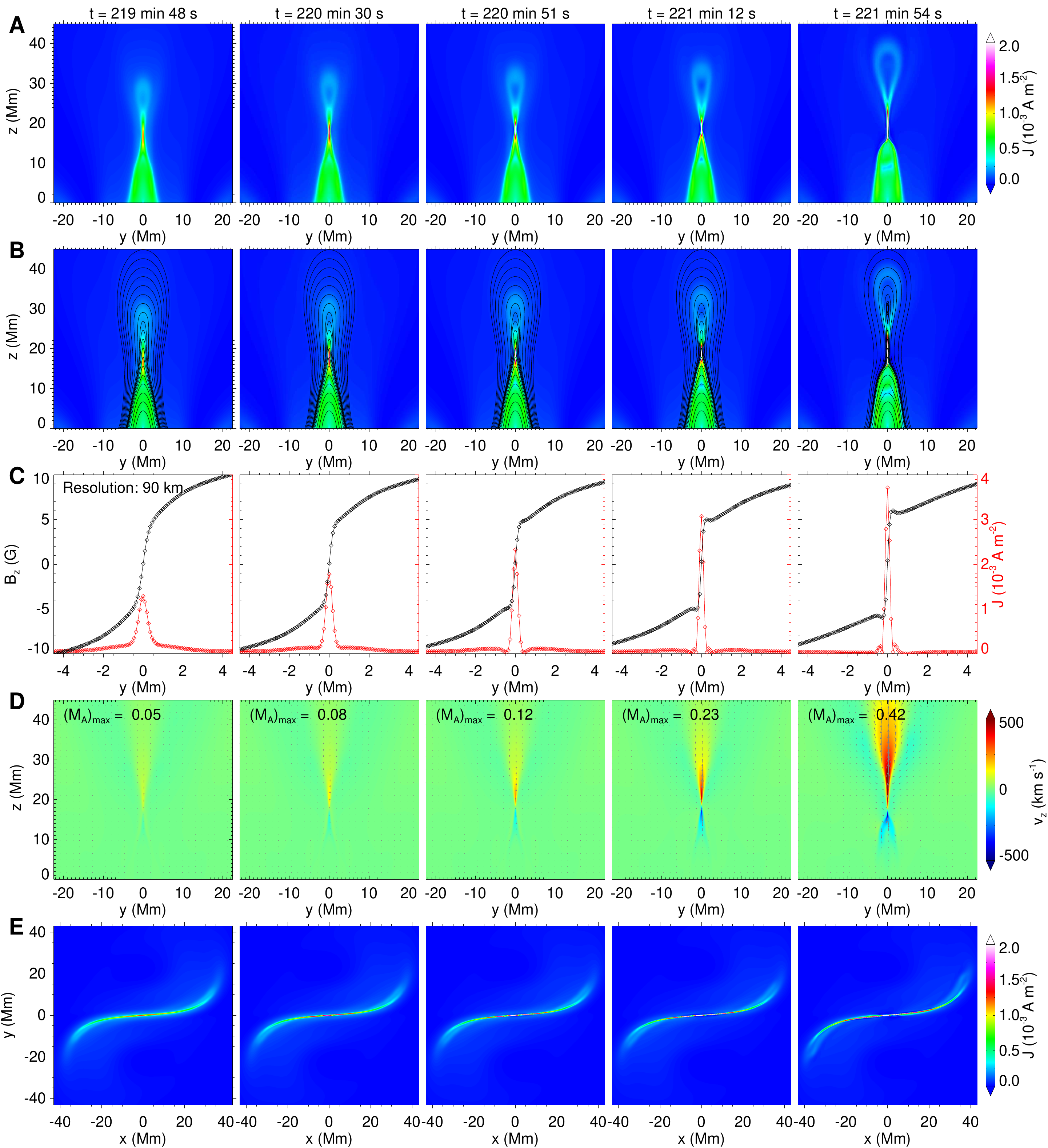}
  \caption{\textbf{Formation of CS and trigger of
      reconnection}. \textbf{(A)} Distribution of current density on
    the central vertical slice, i.e., the $x=0$ slice. \textbf{(B)}
    Projection of magnetic field lines on the same cross section shown
    in \textbf{A}.  \textbf{(C)} 1D
    profile of the magnetic field component $B_{z}$ and current
    density $J$ along a horizontal line crossing perpendicular to the
    CS center (i.e., the point with largest $J$). The diamonds denote
    values on the grid nodes. \textbf{(D)} Distribution of vertical
    velocity on the same cross section shown in \textbf{(A)}. The
    magnitudes of maximal Aflv{\'e}nic Mach number are also
    denoted. \textbf{(E)} Horizontal slice of the current density
    crossing the center of the CS.  }
  \label{F3}
\end{figure*}

% Formation of CS and trigger of reconnection
The key to understanding how the eruption is triggered lies in the
evolution of the central current layer. From Figure~1D to 3A, the
current layer is seen to become progressively thinner, essentially in
its core where the current density is the largest. The thinning of the
current layer occurs in a quasi-static way as driven by the slow
shearing motion applied at the bottom surface
(Methods~\ref{A5}). Eventually, the current layer turns into a CS as
it thins down to the grid resolution, and fast magnetic reconnection
kicks in (Fig.~\ref{F3} shows result for resolution of $90$~km in the
main run; see also results of other runs with higher resolutions in
Methods~\ref{A4} and \Movie~2). This occurs at $221$~min, exactly the
onset time of eruption shown in the energy evolution. The profile of
magnetic field component $B_z$ crossing the CS shows that it thins
down to a tangential discontinuity in numerical sense
(Fig.~\ref{F3}C), since its thickness is only $2 \sim 3$ grid spacings (e.g.,
$0.20$~Mm) which approaches the limit resolvable by the numerical
code. Meanwhile, the current density increases fast in the CS but
decreases elsewhere to nearly zero, rendering its profile crossing the
CS to a Dirac Delta function. This happens when the field, if with
ever increasing magnetic shear in the absence of resistivity,
asymptotically reaches a fully open state \cite{Aly1991,Sturrock1991},
in which the magnetic energy attains its upper limit (i.e., the
Aly-Sturrock limit, which is approximately $1.7\times 10^{30}$~erg in
our case, see Fig.~\ref{F2}A) and all the field is current-free except
in the CS where the current density is infinite
(\EFig~3). However, such open field can never be reached with
finite resistivity, and reconnection is unavoidable once the CS is
sufficiently thin. The reconnection starts with the Petschek
type~\cite{Petschek1964} as its onset is clearly indicated by
bidirectional, collimated Alfv{\'e}nic outflows from the reconnection
site (Fig.~\ref{F3}D). At the onset of the reconnection (or eruption),
the vertical length of the CS is about $10$~Mm, while horizontally it
extends up to over $40$~Mm (Fig.~\ref{F3}E), and thus its aspect
ratios in the vertical and horizontal directions are estimated to be
$\sim 40$ and $150$, respectively. The reconnection quickly reaches a fast rate of $\sim 0.05$ as measured by the inflow Alfv{\'e}nic Mach number, and depends weakly on the Lundquist number (Methods~\ref{A6S}).

\begin{figure*}[htbp]
  \centering
  \includegraphics[width=\textwidth]{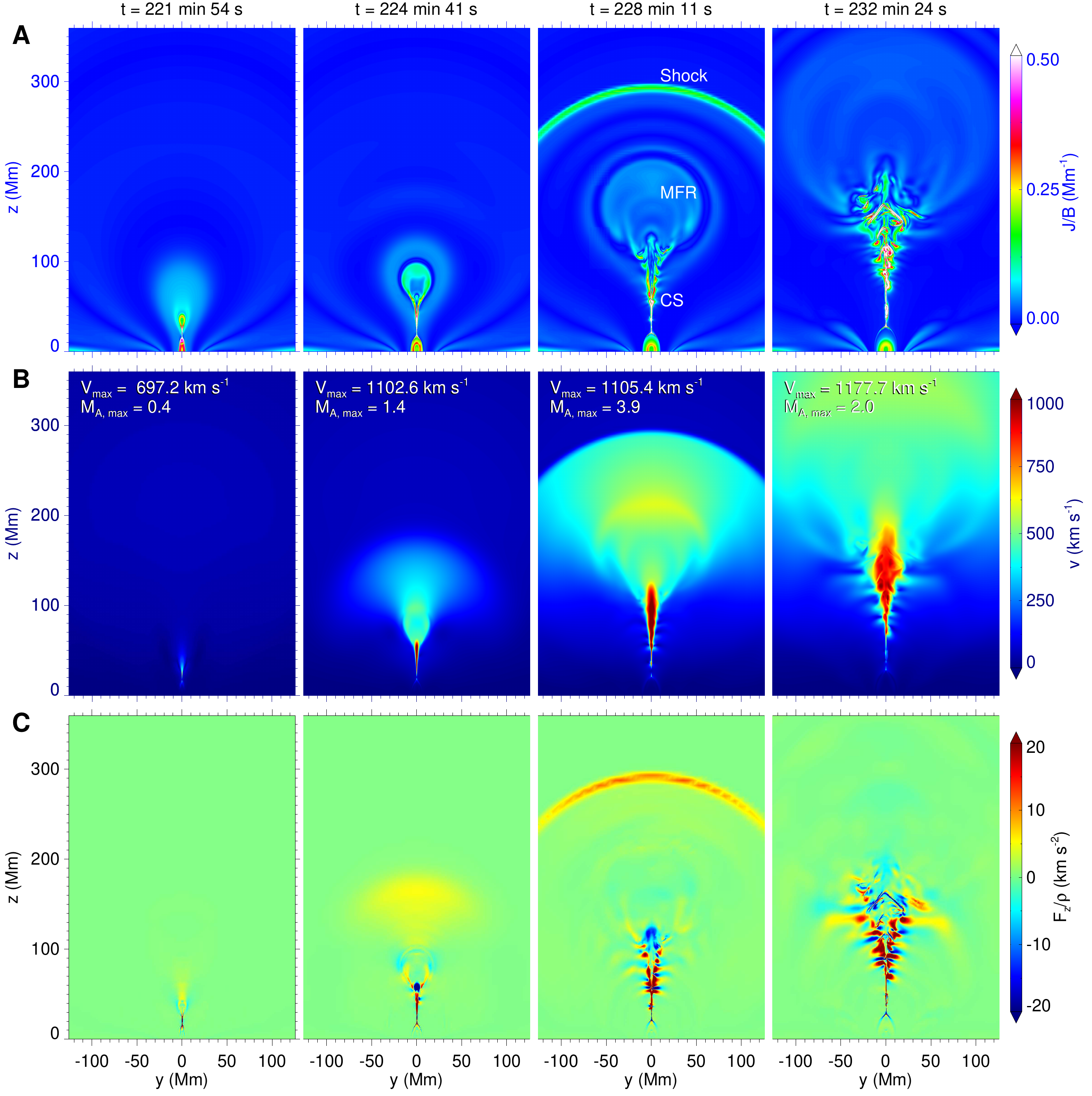}
  \caption{\textbf{Evolution of different parameters during the
      eruption shown in the central vertical slice}.  \textbf{(A)}
    Current density $J$ normalized by magnetic field strength
    $B$. \textbf{(B)} Magnitudes of velocity. The largest velocity and
    Aflv{\'e}nic Mach number are also denoted. \textbf{(C)} The
    vertical component of Lorentz force $F_{z}$ normalized by density
    $\rho$. Also see \Movie~1 for a high-cadence evolution of the
    eruption process.}
  \label{F4}
\end{figure*}

\begin{figure*}[htbp]
  \centering
  \includegraphics[width=\textwidth]{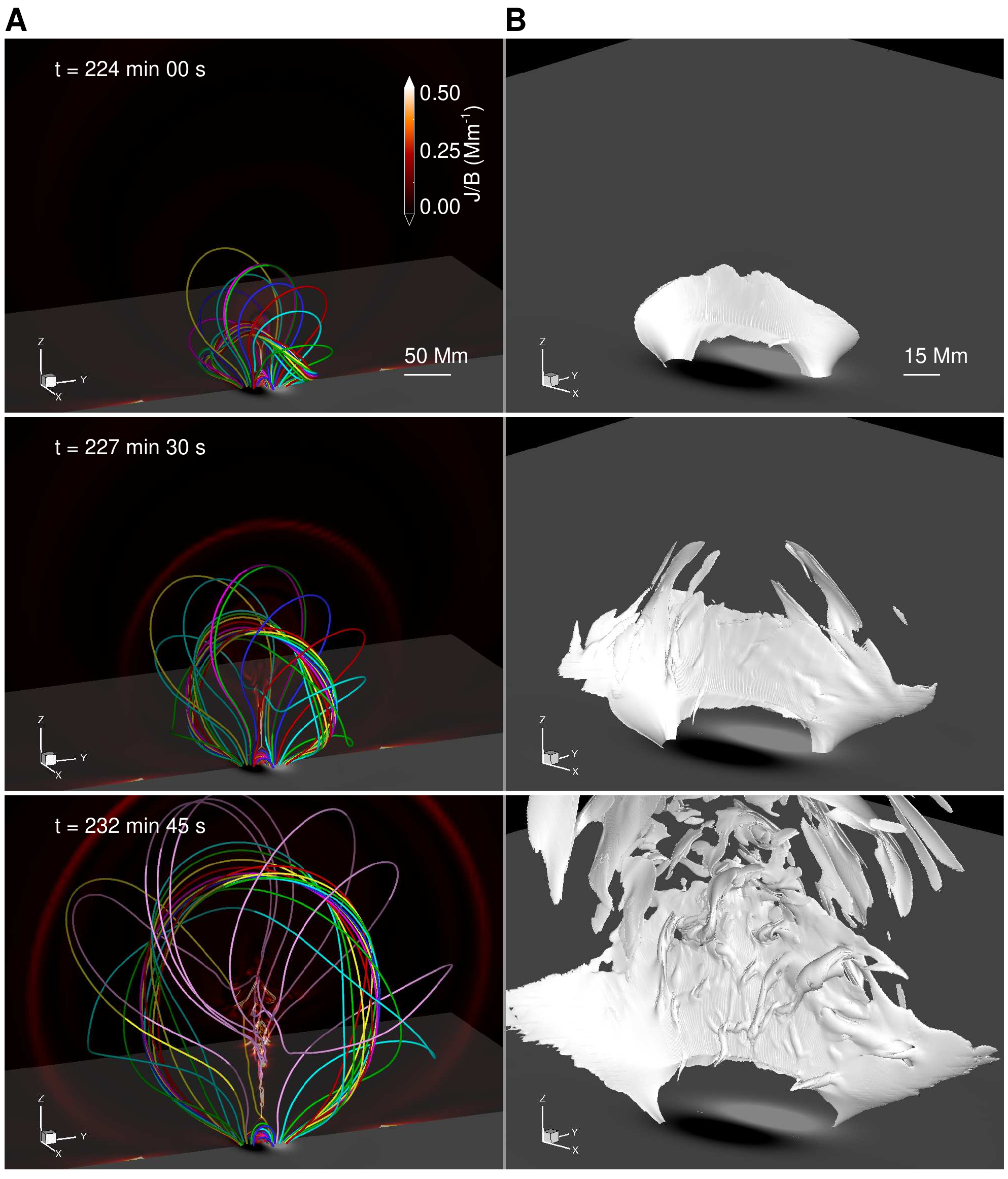}
  \caption{\textbf{Evolution of magnetic field lines and CS
    in 3D during the eruption.}  \textbf{(A)} The magnetic field lines
    are shown by the thick colored lines, and \textbf{the colours are used for a better
    visualization of the different lines}. Note that
    the MFR is weakly twisted in its core but highly twisted in its
    envelope. The bottom surface is shown with distribution of
    magnetic flux. The vertical, transparent slice is shown with
    distribution of current density normalized by magnetic field
    strength, i.e., $J/B$. \textbf{(B)} The CS in 3D configuration is shown
    by iso-surface of $J/B=0.5$~Mm$^{-1}$.}
  \label{F5}
\end{figure*}

Once the reconnection begins, the fast rise of the kinetic energy,
i.e., the eruption, ensues, and the subsequent development of the CS
is extremely dynamical. At large scale, it exhibits a picture of the
standard flare model: a plasmoid originates from the tip of the CS and
rises quickly, leaving behind a cusp structure separating post-flare
loops from un-reconnected fields (Fig.~\ref{F4}A and \Movie~3). The
plasmoid, when initially formed (see Fig.~\ref{F3}B, note that before $t=221$~min 12~s,
the magnetic field exhibits still an arcade configuration, and at $t=221$~min 54~s
a plasmoid is seen at the tip of the CS as a result of reconnection),
is very small with size of a few megameters, and expands
substantially to hundreds of megameters at the end of the
simulation. In 3D, it corresponds to a fast growing MFR (Fig.~\ref{F5}
and \Movie~4), which has a weakly twisted core but wrapped by highly
twisted envelope. %, in line with formation of MFR in the tether-cutting model \cite{Moore2001}.
Meanwhile, an arc-shaped fast magnetosonic
shock forms ahead of the plasmoid and later encloses the whole
erupting structure. All these evolving structures demonstrate a
typical coronal magnetic eruption leading to a CME, as seen in
observations (\Movie~5) as well as previous numerical simulations with
different scenarios \cite{Linker2003,Amari2003A,Torok2018}. With the
magnetic reconnection proceeds continuously, the cusp structure
expands in both vertical and horizontal directions
(\EFig~4). The apex of the cusp ascends with speed of about
22~km~s$^{-1}$. The transverse expansion of the bottom of cusp, which
corresponds to the separation of flare ribbons in observation, goes
with a speed of 11~km~s$^{-1}$, while the leading edge of the CME
reaches a speed of $\sim 600$~km~s$^{-1}$. All these apparent motions
are quantitatively comparable to observed values in typical eruptive
flares \cite{WangH2003,Hinterreiter2018,YanX2018}.

At the onset time of the eruption, the Lundquist number of the CS is
on the order of $10^{4 \sim 5}$ (Methods~\ref{A6}), and thus fast tearing
mode (or plasmoid) instability \citep{Bhattacharjee2009, Huang2010} immediately occurs with the eruption, which fragments
the CS and results in strongly turbulent fluctuations in the
reconnection (Fig.~\ref{F4}). Such small-scale dynamics emerge only in
sufficiently high-resolution computation, and the complexity of the
turbulent reconnecting CS increases in even higher resolutions
(Methods~\ref{A4}). In the later phase of the eruption, the CS
becomes highly fragmented, and many filamentary currents, which are
small plasmoids in two-dimensional (2D) slice and mini flux ropes~\citep{Daughton2011, Nishida2013} in
3D (\EFig~5), are seen in the lower part of the CS
(Fig.~\ref{F5} and \Movie~6). The turbulent reconnection is manifested
as intermittency in the temporal profile of kinetic energy increasing
rate (Fig.~\ref{F2}B) and the distribution and evolution of velocity
(Fig.~\ref{F4}B).

The eruption is powered by magnetic energy and particularly, here the
driver of the eruption comes mainly from the magnetic tension force of
the newly reconnected field through its slingshot effect (Methods~\ref{A9}). As can be
seen in Figure~\ref{F4}C, which shows the vertical component of the
Lorentz force divided by density (i.e., the vertical acceleration of
the plasma), the strongest upward acceleration is always located in
the outflow of the reconnection site, whereas near the central part of
the MFR is mostly downward acceleration. Consequently, as the
reconnected field lines are incorporated in the MFR, they first
experience an impulsive acceleration driven by the strong tension
force, and then slow down quickly in the MFR (\Movie~7). The vertical
velocity is the strongest in the outflow of the reconnection site
(Fig.~\ref{F4}B), and through this high-speed jet flow, the newly
reconnected magnetic field lines continually join and pile up in the
MFR. Thus our analysis suggests that the on-the-fly formed MFR does
not drive the eruption, but is passively pushed by the reconnection
outflow, at least before the CME acceleration reaches its peak (i.e.,
$t=227$~min). Furthermore, at the onset of eruption, the apex of newly
formed MFR is located much lower than the critical height of torus
instability (\EFig~2), i.e., the height at which the decay index
of the strapping field reaches the canonical threshold of $1.5$
\cite{Kliem2006}. Thus, the role of torus instability is minor at the
onset of the eruption.

\section{Discussion}
\label{sec:con}

We have presented a fully 3D MHD simulation of solar eruption produced
in a single bipolar magnetic field, encompassing the entire process
from the gradual accumulation of magnetic free energy to its sudden
release. The simulated initiation process of eruption bears the major
characteristic features of eruptive flares that are associated with
CMEs, such as the formation of coronal sigmoid, the transition from
slow rise to fast acceleration of coronal loop, the elongation and
separation motions of double flare ribbons (Methods~\ref{A8}), the
growth of a flaring cusp structure, as well as the escape and
expansion of a plasmoid, which evolves into a coherent MFR driving a
shock ahead.

Early simulations in 2D or translationally-invariant geometries
\cite{Mikic1994,Choe1996} show that by continuous shearing of its
footpoints a magnetic arcade asymptotically approaches an open state
containing a CS, which is consistent with the Aly-Sturrock
conjecture. However, when one takes into account finite resistivity,
the system experiences a global disruption with reconnection setting
in at the CS, which, in particular, begins at the point with the
largest current density in the CS. %For the first time,
Our simulation
demonstrated this scenario in fully 3D.
%which is the essence of the tether-cutting model.
That is, prior to eruption, the CS forms
internally within the strongly-sheared arcade core in a quasi-static
evolution as driven by photospheric shearing motion. It seamlessly
transforms to a flare CS once reconnection kicks in and the eruption
ensues. As the pre-flare short field lines reconnect to long ones with
a double-arc shape, they are concave upward, thus having strong upward
tension forces which propel upward the newly reconnected flux from the
top of the CS. As a result, more fluxes are allowed to collapse into the
CS and then reconnect, which establishes a positive feedback between
the escape of the newly reconnected flux and the reconnection.
Such a runaway process, as demonstrated being able to be triggered within a sheared, single bipolar field, agrees with the tether-cutting reconnection model~\cite{Moore2001} (which although conjectures that eruption is driven by the unleashed magnetic pressure).
As the reconnection proceeds from the strongly sheared core flux to the
weakly sheared (and nearly current-free) enveloping field, a
large-scale MFR is generated as the core of CME with weakly twisted
axis wrapped by highly winding field lines. Further mediated by the
tearing mode instability, the reconnection runs into a turbulent way,
which strongly fragments the CS.

Compared with other fully 3D simulations
\cite{Linker2003,Amari2003,Aulanier2010} which also start from a
bipolar region that is energized by photospheric flows, ours is unique
in twofold. First, we have shown that the eruption can be initiated by
shearing solely without flux cancellation. Second, it is unnecessary
for an MFR to form before and subsequently trigger the eruption. More
importantly, the fact that the central CS forms in a quasi-static way
distinguishes ours from other 3D simulations of eruption, in which the
CS accounting for flare reconnection forms in a dynamic evolution. For
instance, in the MFR-based models, the CS forms after the rise of the
unstable MFR, which forces the oppositely-directed field lines below
the MFR to approach quickly. Similarly in the breakout model, the
central flare CS forms only after a feedback is triggered between the
expansion of the inner sheared arcade and the breakout reconnection at
the null, in which the central CS is thinned by the fast converging
flow induced by the dynamic expansion of the sheared
arcade~\cite{Karpen2012}. In contrast, the thinning of CS in our
simulation is directly driven by the slow quasi-static shearing, which
requires a sufficiently low numerical diffusion in the computation
and a high accuracy in the line-tied bottom boundary condition, therefore more challenging
than thinning the CS through dynamic inflows (Methods~\ref{A5}).

In summary, our simulation with sufficiently high fidelity demonstrates a fundamental mechanism for solar eruptions triggered and driven by magnetic reconnection, within the simplest magnetic configuration. Whether the mechanism also applies to cases with photospheric flux cancellation as observed mainly in the decaying phase of ARs~\citep{Yardley2018}, or, where a flux rope is built up quasi-statically by much slower reconnection at the photosphere~\citep{Ballegooijen1989} and erupts as it grows to an ideally unstable state \cite{Linker2003,Amari2003,Aulanier2010}, will be examined with comparable high-accuracy simulations.

%for the initiation of solar eruptions, which includes no more ingredient
%than necessary, consistent with the principle of Occam's razor, and can
%potentially be applied universally.

\

\

\noindent \textbf{Correspondence and requests for materials} should be addressed to C.W.J. and X.S.F.

\

\noindent \textbf{Acknowledgments.}  C.W.J. acknowledges support from National Natural Science Foundation of China (NSFC) grants 41822404 and 41731067, the Fundamental Research Funds for the Central Universities (Grant No. HIT.BRETIV.201901), and Shenzhen Technology Project JCYJ20190806142609035.
X.S.F. is supported by NSFC grants 42030204, 41861164026 and 41874202 and the Strategic Priority Program of the Chinese Academy of Sciences, Grant No. XDB41000000.
R.L. is supported by NSFC grants 41774150 and 11925302 and the Strategic Priority Program of the Chinese Academy of Sciences, Grant No. XDB41030100.
X.L.Y. is supported by NSFC grant 11873087, Yunnan Science Foundation for Distinguished Young Scholars under No. 202001AV070004, and the Yunnan Key Science Foundation of China under No. 2018FA001.
Data from observations are courtesy
of NASA SDO and STEREO. The computational work was carried out on
TianHe-1(A), National Supercomputer Center in Tianjin, China.

\

\noindent \textbf{Author contributions.}  C.W.J. conceived the study,
developed the numerical MHD model, performed the result analysis and wrote the
text. X.S.F. contributed to the design of numerical MHD schemes.
R.L., X.L.Y., Q.H., R.L.M., and A.Y.D. contributed
to the result analysis. All authors participated in discussions and revisions on the manuscript.

\

\noindent \textbf{Competing financial interests.}  The authors declare
no competing financial interests.

\

\noindent \textbf{Data availability.}
The data generated by the high-resolution 3D MHD simulations and analyzed for this paper occupy a large amount of approximately 10~TB. Interested parties are invited to contact the corresponding authors to make arrangements for the transfer of those data.

\

\noindent \textbf{Code availability.}  We have opted not to make our numerical code of the MHD simulation publicly available owing to its complexity, which demands expert assistance to set up, run and analyze simulations, and because it is continually being improved and extended, which requires frequent software updates. Interested parties are invited to contact the authors for more detailed information.

%All the data that support the findings of this study are available from the corresponding authors upon reasonable request.
%The code and data that support the findings of this study are available from the corresponding authors upon reasonable request.

\clearpage

%% Figs

\clearpage
\appendix

\section{Methods}

\subsection{Model equations}\label{A1}

We numerically solve the full MHD equations in 3D Cartesian geometry
by the advanced conservation element and solution element (CESE)
method \cite{Jiang2010,Feng2010,Jiang2016NC}. The MHD equations are
given as
\begin{eqnarray}
  \label{eq:MHD}
  \frac{\partial \rho}{\partial t}+\dive (\rho\vec v) = 0,\nonumber
  \\
  \rho\frac{D\mathbf{v}}{D t} = -\grad p+\vec J\times \vec B+\rho\vec
  g + \nabla\cdot(\nu\rho\nabla\mathbf{v}),\nonumber\\
  \frac{\partial \vec B}{\partial t} =
  \grad \times (\vec v \times \vec B - \eta \mu_0 \vec J), \nonumber\\
  \frac{\partial T}{\partial t}+\nabla\cdot (T\vec v) =
  (2-\gamma)T\nabla\cdot\vec v,
\end{eqnarray}
where $\vec J = \nabla \times \vec B/\mu_0$ and $\mu_0$ is the
magnetic permeability in a vacuum. In the momentum equation, a small
kinetic viscosity $\nu$ is used for the purpose of keeping numerical
stability during the very dynamic phase of the simulated
eruptions. Specifically, the coefficient is given depending on the
local spatial resolution $\Delta$ and time step $\Delta t$ as
$\nu = 0.05\Delta^{2}/\Delta t$, which corresponds to grid Reynolds
number of $10$. In the magnetic induction equation, the trigger of
magnetic reconnection depends on the specific choice of magnetic
diffusivity $\eta$. Here we avoid such a sensitivity by relying it
solely on the numerical diffusion, that is, we set $\eta=0$ in the
magnetic induction equation. As such, we can minimize the resistivity
and thus maximize the Lundquist number (with given spatial
resolutions), since any finite value of $\eta$ will result in larger
resistivity than solely the numerical one.  In this sense, magnetic
reconnection occurs when a current layer is sufficiently narrow such
that its width is close to the grid resolution, on which scale the
numerical diffusivity takes effect. We have carefully estimated the
value of numerical diffusivity $\eta_{\rm n}$ (see
Methods~\ref{A6}). In the energy (or temperature) equation, $\gamma$
is the adiabatic index, and here for simplicity as we focus on the
dynamics of the magnetic field, it is set as $\gamma=1$ such that the
energy equation describes an isothermal process.

\subsection{Initial conditions}\label{A2}
We start from a potential magnetic field with vertical component on
the photosphere given by
\begin{equation}
  \label{eq:Bzmap}
  B_{z}(x,y,0) = B_{0}e^{-x^{2}/\sigma_{x}^{2}}(
  e^{-(y-y_{c})^{2}/\sigma_{y}^{2}}-
  e^{-(y+y_{c})^{2}/\sigma_{y}^{2}}),
\end{equation}
where $B_{0} = 37.2$~G, $\sigma_{x} = 28.8$~Mm,
$\sigma_{y}=\sigma_{x}/2$, and $y_{c}=11.5$~Mm. Such magnetic
configuration is similar to that used in reference
\cite{Amari2003}. As shown in \EFig~1A, it mimics a bipolar
solar AR of typical size, but the magnetic field strength is weaker
than that of real sunspots by a least one order of magnitude to avoid
a too heavy burden on computation.

The background atmosphere is stratified by solar gravity with density
$\rho=2.3\times 10^{-15}$~g~cm$^{-3}$ at the bottom and an uniform
temperature of $T=10^6$~K (corresponding to sound speed of
$110$~km~s$^{-1}$). Here we find that if using the real number of the
solar gravity ($g_{\odot} = 274$~m~s$^{-2}$), it results in a pressure
scale height of $H_{p} = 43.8$~Mm, by which the plasma pressure and
density decay with height much slower than the magnetic field. With
the weak magnetic field strength we used, the plasma $\beta$ will
increase with height very fast to above $1$, which is not realistic in
the low corona. To make the pressure (and density) decrease faster in
the lower corona, we modified the gravity by defining it as
\begin{equation}
  \label{eq:gravity}
  g = \frac{k}{(1 + z/L)^2}g_{\odot}.
\end{equation}
where $k=5.7$ and $L=76.8$~Mm.  By this, we get a plasma $\beta<1$
mainly within $z <120)$~Mm and the smallest value is $4\times 10^{-3}$
(\EFig~1C). Note that with the energizing of the magnetic field
by the shearing flow, the magnetic field inflates and its strength
increases significantly in the upper volume, and the plasma $\beta$
will decrease further. As can be seen in \EFig~1, just prior to
the eruption, the plasma $\beta$ is much smaller than unity in the
height up to 200~Mm, and the Alfv{\'e}n speed higher than
$1000$~km~s$^{-1}$ within $z < 100$~Mm.

\subsection{Boundary conditions}\label{A3}
\label{sec:init}

On the bottom boundary ($z=0$), we apply the surface rotation flow
(with $v_{z} = 0$) to add free magnetic energy to the initial
potential field. To ensure that such flow will not modify the magnetic
flux distribution $B_{z}$ at the photosphere, the flow is
incompressible and the streamlines coincide with the contour lines of
$B_{z}$. Specifically, the surface velocity is set as
\begin{equation}
  v_{x} = \frac{\partial \psi(B_{z})}{\partial y};
  v_{y} = -\frac{\partial \psi(B_{z})}{\partial x};
\end{equation}
with $\psi$ given by
\begin{equation}
  \psi = v_{0}B_{z}^{2}e^{-(B_z^2-B_{z, \rm max}^2)/B_{z, \rm max}^2 }
\end{equation}
where $B_{z, \rm max}$ is the largest value of the photosphere
$B_{z}$, and $v_{0}$ is a constant for scale such that the maximum of
the surface velocity is $4.4$~km~s$^{-1}$, close to the magnitude of
typical flow speed in the photosphere ($\sim 1$~km~s$^{-1}$). The flow
pattern is shown in \EFig~1A and B. As the flow speed is smaller
than the sound speed by two orders of magnitude and the local
Alfv{\'e}n speed by three orders, it stresses the corona magnetic
field very slowly. Such flow mimics the frequently-observed sunspot
rotation during evolution of ARs \cite{Brown2003,YanX2007,YanX2012},
and similar rotational flows have be employed in numerous numerical
simulations for the same purpose of energizing pre-eruption fields but
the magnitude is often larger than ours by an order
\cite{Amari1996B,Tokman2002,Torok2003,DeVore2008}. The energizing
phase is linearly ramped on by a time of $t=10.5$~s.

We fix plasma density on the bottom surface as being their initial,
uniform value, because the surface flow is
incompressible. Furthermore, as the velocity is prescribed there and
$B_z$ does not change, we only need to specify how the horizontal
magnetic field evolves. To deal with this, many codes use simple
linear extrapolation from the inner grid points, which, however, could
falsely increase the magnetic energy (as shown in \SFig~1). To
minimize numerical errors introduced by any inappropriate treatment of
the boundary conditions, we directly solve the ideal induction
equation
\begin{equation}
\label{photo_B_equ}
  \frac{\partial \vec B}{\partial t} =
  \grad \times (\vec v \times \vec B)
\end{equation}
on the bottom surface. By this we can self-consistently update the
magnetic field and simulate the line-tied effect at the photosphere,
which is essential for the success of the simulations in this
type. Solving this equation is realized by second-order difference in
space and forward difference in time. Specifically, on the bottom
boundary (we do not use any ghost or guard cell), we first compute
$\vec v\times \vec B$, and then use central difference in horizontal
direction and one-sided difference (also 2nd order) in the vertical
direction to compute the convection term
$\grad \times (\vec v \times \vec B)$. We have further checked the
accuracy with which the line-tied condition is implemented. To
illustrate this, we traced the successive movements of footpoints of a
single field line that is sheared through the area with the largest
surface flow, which is shown in \SFig~2. Since both footpoints
of the field line convect with surface flow, we assume that the
footpoint in the positive polarity (i.e., the $y>0$ part) moves with
flow exactly and trace the field line to its conjugated footpoint. If
the line-tied condition is accurately implemented, the positions of
the conjugated footpoint should be exactly the positions expected due
to the applied flow. As shown in \SFig~2, we traced the movement
of the footpoints with a time cadence of $35$~min until the onset of
the eruption. As can be seen, the exact footpoints are almost
excellently matched by the computed ones, except that the last one
(i.e., $t=6\times 35$~min) has a finite, but still small, offset of
about $0.5$~Mm. Such a finite offset is more likely resulted by
numerical errors in magnetic field line tracing rather than an actual
slippage.  In particular, as the field immediately prior to the
eruption is strongly sheared, it forms a quasi-separatrix layer (QSL)
in which the gradient of field-line mapping is extremely large, and
small errors in field-line tracing can result in large offset. We note
that a sufficient resolution of the driving surface is also important
to avoid unwanted slipping of the field lines (see Methods~\ref{A7}).

On the side and top boundaries, we fixed the plasma density,
temperature and velocity. The tangential components of magnetic field
are linearly extrapolated from the inner points, while the normal
component is modified according to the divergence-free condition to
avoid accumulation of numerical magnetic divergence near the
boundaries. Furthermore, the simulation runs are stopped before the
disturbance by the eruption reaches any of these boundaries to
minimize the influence of these numerical boundaries on the
computation.

It is worth noting that the time scale of the quasi-static,
pre-eruption evolution is determined by the speed of the surface
motion. Since the surface velocity we used is still a few times (say,
5) larger in magnitude than typical photosphere flow, our simulated
pre-flare evolution time, if compared with the realistic time scale,
should be multiplied by a factor of 5, thus corresponding to roughly
one day. On the other hand, the time scale of the eruption is
controlled by the evolution in the coronal volume, and thus is not
changed.

\subsection{Grid setting and influence of eruption onset by different resolutions}\label{A4}
The computational volume spans a sufficiently large box of
approximately $(-370, -370, 0)$~Mm $< (x, y, z) < (370, 370, 740)$~Mm
(where $z=0$ represents the solar surface) such that during eruption
the saturation of kinetic energy (i.e., the completion of CME
acceleration) occurs prior to the moment when any disturbance reaching
the side and top boundaries. The full volume is resolved by a
block-structured grid with adaptive mesh refinement (AMR). The AMR is
designed to automatically resolve with highest resolutions the narrow
layers with strong currents (mainly in the low $\beta$ region) as well
as the regions where the magnetic field has a strong gradient (see
\Movie~8).  Specifically, the base resolution is
$\Delta x = \Delta y = \Delta z = \Delta =2.88$~Mm, and we carried out
four different runs with different highest resolutions, including
$\Delta = 180$~km, $90$~km, $45$~km and $22.5$~km (will be referred to
RES0, RES1, RES2, and RES3 respectively, where RES1 is the main
run). During the calculation any location with $\JB> 0.1$ will be
refined to the highest resolution, and any location with strong
magnetic field gradient or strong current, with criteria given by
$|\nabla (B^{2}/2)|\Delta /\rho > 10 $ and
$|(\vec B\cdot\nabla)\vec B|\Delta /\rho > 10$ respectively, will also
be refined. Furthermore, at the bottom boundary where the surface flow
is applied, the resolution is forced to be no less than $180$~km. If
using uniform grids, to reach these resolutions requires grid numbers
of $4096^3$, $8192^{3}$, $16384^3$, and $32768^{3}$, respectively,
which is formidable.

The energy evolution curves from the different runs are compared in
\SFig~3. In all the runs, the energies evolve similarly, with
pre-eruption to eruption clearly denoted by sharp transitions of
energy curves and their changing rates. Note that the different
resolutions do not change the evolution in the pre-eruption phase,
i.e., the ideal MHD process, indicating that our simulation converges
in the ideal MHD regime. However, the onset time of eruption in the
runs is clearly different; it is postponed by about $100$~s
incrementally from run of RES0 to RES3. This is because, as we have
mentioned before, the reconnection results from numerical diffusion
when the pre-eruption CS is thin enough as close to the grid
resolution. In this sense, the onset time of reconnection depends on
the grid spacing, since with a smaller grid size a thinner CS can
develop, and can sustain stronger current density (and thus more free
energy), which needs more time to accumulate, and thus the onset of
reconnection in the CS is postponed relative to runs with lower
resolutions. This effect is exactly shown in \SFig~3 and
\Movie~2. With the resolution increased from RES0 to RES3, the
thickness of the CS at the eruption onset decreases from approximately
$500$~km to $50$~km, and correspondingly, the peak current density in
the CS increases proportionally from approximately
$2\times 10^{-3}$~A~m$^{-2}$ to $14 \times 10^{-3}$~A~m$^{-2}$.
We note that there is a clear link between the field at the eruption onset and the corresponding open-field
configuration by comparing the peak current density and the maximal current density
of the open field discretized with the same grid resolution.
For instance, as can be seen in \EFig~3 which shows the open field
discretized with resolution of $90$~km, the current density in the CS is not infinite
but rather changes with height and at approximately $z=10$~Mm it reaches the maximum of
$13\times 10^{-3}$~A~m$^{-2}$. This value is close to triple of
the peak current density ($\sim 4\times 10^{-3}$~A~m$^{-2}$) in the CS of RES1 run, which is
consistent with the fact that the CS thickness in RES1 run is
approximately triple of $90$~km. The
fast reconnection always starts when the CS thins down to about $2 \sim 3$
grid spacings, independent of how small the grid spacing is, and once
the reconnection kicks in, the eruption starts. These tests with
different resolutions confirm again that reconnection plays the key
role in triggering the eruption. It is expected that by using further
higher resolutions the CS will be thinner and the current density
can increase accordingly. If needed (though formidable in computation), it could be
even thinned down to the kinetic scale, i.e., the ion gyro-radius or
inertial length, where the MHD approximation fails, as the current density may
exceed the threshold of a microscopic instability and anomalous
resistivity arises to trigger reconnection~\citep{Shibata2001}.
% rather than ideal MHD instability. in real world, the current is
% large enough, resistivity or microscopic instability will be
% triggered

Once the eruption is triggered, it evolves rather differently on small
scales in the different runs, since with the higher resolution, the
earlier the tearing instability can be triggered, and the more complex
the turbulent reconnecting CS will be (\Movie~9), and thus there are
more fluctuations seen in the curves of energy evolutions
(\SFig~3B and C).

%\section{Supplemental Text}

\subsection{Quasi-static formation of CS}\label{A5}
A key aspect of our 3D simulation is that the CS forms in a
quasi-static way. Prior to the eruption onset, the horizontal flow on
two sides of the current layer converges to it and plays the role of
thinning the current layer to CS (\Movie~2C and F). As the converging
motion is directly driven by the surface shearing motion, the
converging speed, i.e., the thinning speed of the current layer, is on
the same order of the surface flow speed, which is several km~s$^{-1}$
(for instance, see \SFig~4 for the runs of RES2 and RES3). This
is distinguished from other 3D simulations of eruption where the
central CS accounting for the flare reconnection is formed in a
dynamic way. For example, in the MFR-based models, the CS forms after
the rise of the erupting MFR, which forces the oppositely directed
field lines below the MFR to approach each other quickly. Similarly,
in the breakout model, firstly the reconnection begins at the null and
triggers a feedback between the expansion of the inner sheared arcades
and the breakout reconnection, making the system run into a dynamic
phase, and then the main flare CS forms.

In numerical simulations, it is much more challenging to form a CS in
a quasi-static way than in a dynamic way (especially in the absence of
magnetic topological separatries such as the magnetic null
point). This is because, a current layer will diffuse and broaden by
the numerical diffusion $\eta_{\rm n}$.  Say, for a thin current layer
with thickness of $l$, it will diffuse (or broaden) with a speed of
$v_{\rm d} = \eta_{\rm n}/l$ in the absence of inflow
\cite{Priest1987Book}. The current layer can only be thinned by a
inflow $v_i$ larger than the diffusing speed, i.e., $v_i > v_{\rm d}$.
This means to form a CS with thickness $l$, one must have
$\eta_{\rm n} < v_i l$. In the dynamic formation of CS, this can be
easily fulfilled since the inflow speed $v_i$ is sufficiently large to
exceed the diffusion speed $v_{\rm d}$. In our simulation, the inflow
$v_i$ is as slow as several km~s$^{-1}$.  For instance, in our RES3
run, $l \approx 60$~km, $v_i \approx 3$~km~s$^{-1}$, which requires
the numerical diffusivity
$\eta_{\rm n} < v_i l \approx 180$~km$^2$~s$^{-1}$. Thus only with a
sufficiently small numerical diffusivity can the CS form in a
quasi-static way. Note that the quasi-static formation of CS is
directly related to degree of the magnetic shear and free energy
storage. If the numerical diffusivity is too large, the CS cannot form
since magnetic free energy diffuses faster than the injection rate
from the bottom surface. Such a numerical effect is also seen in a
previous 2D simulation~\cite{Shiota2008}, and this might be the key
reason why early simulations fail to reproduce the runaway tether-cutting
model in fully 3D since they have too large numerical diffusion to
form a thin CS in the pre-eruption quasi-static evolution.  As a
result, the numerical resistivity takes effect much earlier before the
CS forms, and the resulted reconnection, which is slow, will readily
build up MFR in the pre-eruption phase. Such an MFR formation process,
aided with surface converging motion (or flux cancellation), has been
commonly seen in earlier 3D simulations
\cite{Amari2003,Aulanier2010}. This also explains why many 3D
numerical simulations tend to support the MFR-based eruption
scenarios. However, we note that the quasi-static formation of CS is a
key process in the pre-eruption evolution, and besides, it is also
essential for explaining confined flares, in which only flare emission
is observed but without noticeable movement of coronal loops
\cite{Jiang2016ApJ}, indicating that CS thinning and the subsequent
reconnection are not driven by dynamic evolution.

\subsection{Estimation of effective magnetic diffusivity and Lundquist number}\label{A6}
In the solar corona, the magnetic diffusivity as derived from the
Spitzer resistivity~\cite{Spitzer1962} is
$\eta \approx 1$~m$^{2}$~s$^{-1}$, and at a typical length scale,
e.g., $L = 10$~Mm, and Alfv{\'e}n speed of
$ v_{\rm A} = 10^3$~km~s$^{-1}$, the coronal environment has extremely
large Lundquist number of $S = L v_{\rm A}/\eta \approx 10^{13}$. Thus
fast reconnection only occurs when a CS is sufficiently thin, e.g.,
with thickness of a few to tens of meters. Although it is prohibitive
to reach such a high Lundquist number in numerical simulations of
large-scale eruptions, we attempted to mimic the coronal conditions as
much as we can. Here we estimate the values of the numerical
diffusivity $\eta_{\rm n}$ and the corresponding Lundquist number in
the runs with different resolutions. Since the eruption is triggered
by reconnection, the different values of actual diffusivity will
naturally result in different onset times of reconnection and thus
eruption. That is, the larger the diffusivity is, the earlier the
eruption is triggered. As our MHD code has a second-order accuracy,
the numerical diffusivity decreases with resolution by a power of two,
i.e., $\eta_{\rm n} \propto \Delta^{2}$. For example, in RES0 the
numerical diffusivity is four times of that in RES1. If the onset time
of eruption in RES1 using a particular value of $\eta$ coincides with
the eruption onset time in RES0 without explicit diffusivity, then we
can estimate that $\eta = 3\eta_{\rm n}$ in RES1, as such the total
diffusivity in RES1 is $4\eta_{\rm n}$, which equals to the numerical
diffusivity in RES0.

Thus, to quantify the $\eta_{\rm n}$ in RES1, we run a series of tests
with RES1 using different values of explicit diffusivity $\eta$. Since
this needs many runs, we speedup the bottom boundary flow by 5 times
such that the largest speed is $22$~km~s$^{-1}$ to save the computing
time. By this speed, it is still a slow quasi-static driving of the
coronal field, though the pre-eruption evolution is faster than that
in our main run. \SFig~5A shows the energy curves for the run
with $\eta =0$, which is very similar to those of the main run shown
in \Fig~\ref{F2}, except that the timing is different and the
pre-eruption kinetic energy is somewhat larger.  Then in
\SFig~5B, we compare the RES1 using a sequence of $\eta$ with
the RES0 using $\eta=0$ by plotting the changing rate of kinetic
energy from the pre-eruption to eruption phases. Clearly, the larger
the diffusivity is, the earlier the eruption is triggered.  From these
tests, we can estimate the numerical diffusivity with RES1 to be
$\eta_{\rm n} \approx 1.3 \times 10^{-3} $~Mm$^{2}$~s$^{-1}$ or
$1300$~km$^{2}$~s$^{-1}$, and thus the Lundquist number
$S = L v_{\rm A}/\eta_{\rm n} \approx 1\times 10^{4}$, where $L=10$~Mm
is the vertical length scale of the CS immediately prior to eruptions
and $v_{\rm A} = 10^{3}$~km~s$^{-1}$ is the typical Alfv{\'e}n speed
around the CS. Accordingly, the numerical diffusivity in runs of RES2
and RES3 are approximately $300$~km$^{2}$~s$^{-1}$ and
$80$~km$^{2}$~s$^{-1}$ (the latter is consistent with aforementioned
requirement for CS formation in Methods~\ref{A5}, i.e.,
$\eta_{\rm n} < v_i l = 180$~km$^2$~s$^{-1}$ ), and the Lundquist
numbers are $4\times 10^{4}$ and $1.6\times 10^{5}$, respectively. The
Lundquist numbers further increase with the length of the CS, which
grows as the eruption goes on. We note that the Lundquist numbers in
our fully 3D simulations are comparable to and even larger than values
used in 2D simulations of the similar type
\cite{Mikic1994,Choe1996,Shiota2008}. If considering to extrapolate
our values to the real coronal conditions, we would needs a grid
resolution of $\Delta \approx \Delta_{\rm RES3}/10^4$, on which the
effective diffusivity can then be
$\eta = \eta_{\rm RES3}/10^8 \approx 1$~m$^{2}$~s$^{-1}$, the
Lundquist number $S = S_{\rm RES3} \times 10^8 \approx 10^{13}$ and
the CS thickness $l \sim 3\Delta \approx 7$~m, which are all
consistent with the classical values derived from the Spitzer
resistivity.

\subsection{Magnetic reconnection rate and indication for fast reconnection}\label{A6S}
In \SFig~6A, we compare the reconnection rates, measured by the inflow Alfv{\'e}nic Mach number, of the 4 runs with increasing resolution (or Lundquist number). As the highest resolution run (RES3) stops earlier when the reconnection lasts for about 3 minutes, we thus compare the reconnection rates averaged in the initial 3 minutes for all the runs. All the runs show fast reconnection rate of $\sim 0.05$, which is weakly dependent on the Lundquist number. From RES0 to RES2, the reconnection rate decreases slightly while for RES3 it increases a little bit again.

The reconnection in our simulations starts with the Petschek type, i.e., with onset at a single X-point. This is mainly because the numerical resistivity in our code actually mimics an anomalous resistivity since it is almost negligible in the smooth field region and becomes effective (or turns on) only when the current density exceeds a critical value such that the thickness of CS is close to the grid size. It is similar to using an explicit form of resistivity that depends sensitively on the local current density~\citep{Yokoyama1994}, which can lead to the Petschek-type reconnection. Furthermore, the CS is formed with very non-uniform current density and thickness along its length (\SFig~7), i.e., during the formation of CS, the current density grows much faster at its peak-value point than elsewhere, and thus the reconnection should be first triggered at the point with the largest current density. Such inhomogeneity is inherent to the 3D nature of the magnetic configuration.

Our scaling experiments with the 4 different resolutions indicate that the CS can be further thinned (especially at the point of peak value of current density) with higher resolutions, even possibly down to the scale at which the micro-instability is triggered and creates true anomalous resistivity for fast Petschek reconnection~\citep{Shibata2001}. But before this, the Lundquist number of the CS should be well above the critical value ($\sim 10^4$) for the nonlinear plasmoid instability, which will also be triggered and realize again fast reconnection nearly independent of Lundquist number~\citep{Bhattacharjee2009, Huang2010}. Our simulation indicates this clearly since the plasmoids emerge earlier and their number grows faster in the runs with higher resolutions (\Movie~9). Additionally, the turbulence as excited by the plasmoid-mediated reconnection, can also enhance the reconnection rate~\citep{Lazarian1999,Kowal2009}. \SFig~6B and C show the evolution of the maximum value of velocity in each run, and the comparison of different resolutions show that the impulsiveness of plasma acceleration increases and the turbulent fluctuation starts sooner as the resolution increases. Especially in RES3, the turbulence is induced within nearly 1 minute after the reconnection starts, and this explains partially why the reconnection rate of RES3 exceeds RES2.

Since all these mechanisms as shown in our simulation, namely the Petscheck-type reconnection, the plasmoid instability as well as the induced turbulence, can produce fast reconnection with rate scaling weakly on Lundquist number, there is a strong indication of fast reconnection with much higher resolutions, or equivalently, much higher Lundquist numbers, eventually reaching the realistic values of $\sim 10^{13}$ in the corona. However, the very details on how reconnection works at such high Lundquist number and small scales are unknown as it is prohibitive to simulate in current computations, and moreover they are related to the complex coupling between the MHD and kinetic scales, which is beyond the scope of this work.

\subsection{A ``failed'' simulation with inaccurate boundary conditions}\label{A7}
As aforementioned, a sufficiently high resolution at the bottom
boundary is also required to accurately implement the line-tied
boundary condition, and further to form sufficiently thin CS. To show
this, we run an experiment with the highest resolution of $90$~km
(same as RES1) to resolve the current layer but using a four times
lower resolution (i.e., $720$~km with respect to $180$~km in the main
run) at the bottom boundary.  As can be seen in \SFig~8A, this
run failed to produce an eruption, as no impulsive release of magnetic
energy accompanied with fast rise of kinetic energy is seen. In
contrast to the high-resolution run, there is a considerable amount of
magnetic energy loss in the surface driving process. This is because
the line-tying condition is not accurately achieved (see
\SFig~8B) in the lower resolution due to numerical
errors. Consequently, the field lines slip backward and are unable to
obtain the amount of shear as needed for the CS formation. If putting
the two negative factors together, i.e., the MHD code has a large
numerical diffusion and the line-tied boundary condition is not
accurately implemented, it is very likely that the magnetic energy
eventually saturates once the net energy injection rate from the
bottom surface is totally cancelled by the diffusion rate. Therefore
eruption can not happen even if the surface shearing is continuously
applied.

\subsection{Evolution of magnetic topology and formation of MFR}\label{A8}
The magnetic topology is a simple arcade until the eruption onset and,
once reconnection begins, it transforms to complex one having a highly
twisted MFR formed during the eruption. To reveal the variation of
magnetic topology, we inspected the distribution and evolution of two
parameters, the magnetic squashing degree and magnetic twist
number, which are commonly used for study of 3D magnetic fields and their
dynamics~\citep{Aulanier2012, Janvier2013,Inoue2013,Savcheva2016, LiuR2016, DuanA2019}.
The magnetic squashing degree $Q$ quantifies the gradient of
magnetic field-line mapping with respect to their footpoints, and it
is helpful for searching QSLs of magnetic fields~\cite{Titov2002},
which can have extremely large values of $Q$ (e.g., $\geq 10^5$) and
are preferential sites of magnetic reconnection. Specifically, for a
field line starting at one footpoint $(x,y)$ and ending at the other
footpoint $(X, Y)$ where $X$ and $Y$ are both functions of $x$ and
$y$, the squashing degree $Q$ associated with this field line is given
by~\cite{Titov2002}
\begin{equation}
  \label{eq:Q}
  Q = \frac{a^{2}+b^{2}+c^{2}+d^{2}}{|ad-bc|}
\end{equation}
where
\begin{equation}
  a = \frac{\partial X}{\partial x},\ \
  b = \frac{\partial X}{\partial y},\ \
  c = \frac{\partial Y}{\partial x},\ \
  d = \frac{\partial Y}{\partial y}.
\end{equation}
The magnetic twist number $T_w$~\cite{Berger2006} is defined for a
given (closed) field line by taking integration of
$T_w=\int_L \vec J\cdot \vec B/B^2 dl /(4\pi)$ along the length $L$ of
the field line between two conjugated footpoints on the
photosphere. Note that $T_w$ is not identical to the classic winding
number of field lines about a common axis, but an
approximation of the number of turns that two infinitesimally close
field lines wind about each other~\cite{LiuR2016}.

\SFig~9 and \Movie~10 show evolution of magnetic squashing
degree $Q$ on the central vertical slice, as well as both the $Q$,
$T_w$ and current density $J$ on the bottom surface, respectively. As
can be seen, the formation of CS is accompanied by the formation of
corresponding QSLs. Initially the distribution of $Q$ is rather
smooth, and with narrowing of the current layer, there is an evident
increasing of $Q$ (reaching $\sim 10^5$) in two thin strips of J
shape, i.e., QSLs, on either side of the PIL on the bottom
surface. These QSLs correspond to the intersection of the CS with the
photosphere, and thus two thin ribbons of enhanced current density are
also seen, co-spatial with the QSLs. With onset of the reconnection,
the two J-shaped QSLs and the current ribbons on the bottom surface
evolve rapidly, which corresponds to the apparent motion of footpoints
of the field lines that were undergoing reconnection, or simply the
motion of observed flare
ribbons~\cite{QiuJ2002,Savcheva2016,JiangC2018}.

In the early phase, the reconnection is fully a 3D manner with a
strong guide field component (i.e., $B_x$) because joining in the
reconnection is mainly the strongly-sheared, low-lying flux. While in
the later phase, it transfers into a quasi-2D manner, which consumes
mainly the large-scale, overlying flux that is weakly sheared.  Such
transition is clearly manifested by the fast elongation of the arms of
the J-shaped QSLs along the direction of the PIL on the bottom
surface, which agrees with observed elongation of flare
ribbons~\cite{QiuJ2017}. Specifically, the arm of the QSL in the north
(south) of the PIL spreads to the left (right), and as a result, the
observed two ribbons naturally exhibit an evolution pattern of
strong-to-weak shear~\cite{SuY2007}. As the eruption proceeds, more
and more magnetic fluxes reconnect, and consequently, the two J-shaped
QSLs continuously separate with each other, in agreement with the
well-known separation motion of two flare ribbons. At the end of the
simulation, they have swept to near the center of each magnetic
polarity (or the umbra of the sunspots).  On the central vertical
slice, the QSLs intersect with each other, developing into an X shape,
which is referred to a hyperbolic flux tube (HFT)~\cite{Titov2002},
and the intersection X point is essentially the main reconnection site
(in analogy to the null point in a 2D X-shaped reconnection
configuration). As the reconnection proceeds, the X point of the HFT
rises upward progressively with the cusp region below expanding.

Starting from the hooks of the J-shaped QSLs, twisted magnetic flux
(as indicated with $T_w < -2$) begins to form owing to the tether-cutting
reconnection, which creates long field lines connecting the far ends
of the two pre-reconnection sheared field lines. With the twisted flux
accumulated through the continuation of reconnection, the areas
occupied by the footpoints of the highly twisted field lines at the
hooks expand. Consequently, the hook of each J-shaped QSL continuously
extends inward until it reaches the arm, forming a closed curve
encircling the highly twisted flux (see the panels of
$t=226$~min~$06$~s). Such a transition of QSLs ought to be observed as
flare ribbons that gradually forms close
rings~\cite{WangW2017}. Accordingly, as can be seen in the vertical
cross section, the QSLs form a closed tear-drop shape connecting the
HFT. Thus, in magnetic topology, at this moment the MFR is fully separated with its
surrounding magnetic field by the
QSLs~\cite{Demoulin1996JGR,Savcheva2012a,Janvier2014}. Interestingly,
this moment is close to the peak time of the
magnetic energy release rate and the kinetic energy increasing
rate. We note that the distribution of magnetic twist degree is rather
inhomogeneous, and the most strongly twisted flux is seen around the
boundary of the MFR, meaning that the newly-formed MFR has weaker
twisted axis wrapped by stronger winding field lines. It is also
worthy noting that the details of the QSLs associated with the MFR
become extremely complex in the later phase with many fast evolving
small structures (see \Movie~10) because of the turbulence excited in
the reconnection. The current ribbons also show small evolving
kernels, which might correspond to the observed bright kernels or
knots within flare ribbons~\cite{JingJ2016}.

\subsection{Analysis for the driver of the eruption}\label{A9}
Once the MFR is formed, it can be subject to the torus instability if its axis reaches a height where the external (or strapping) magnetic field decays fast enough with height such that the hoop force of the MFR exceeds the strapping force in its subsequent expansion. Thus, the torus instability can occur regardless of whether the MFR is formed in equilibrium or dynamically. To check whether the MFR formed in our simulation is affected by such instability in its acceleration, we calculate the decay index of the external field at the time when the MFR first forms, which is $t=221$~min~33~s (in RES1).
Since it is only $B_y$ that actually straps the MFR, the decay index is defined as $n=-d\ln (-B_y)/d\ln(z)$.
Here the external field $B_y$ consists of two components by two different sources of currents: one is the field generated by current below the bottom surface, i.e., the potential field; the other is the field generated by current above the surface but below the MFR, which is mainly the CS. The reason why the latter should be considered is that, as the current in the CS follows in the same direction as that of the MFR, it attracts the MFR and thus prevents the MFR's outward expansion by enhancing the external field overlying the MFR. However, it is difficult to separate the current in the CS with that of the MFR as the coronal current distributes continuously from the CS to the MFR. To give a reasonable approximation of the external field, we use the total magnetic field $B_y$ at an earlier time ($t=217$~min) than the eruption onset, since at that moment the MFR is not yet formed (thus there is no current of MFR), and meanwhile the current in the corona is close to that in the CS below the MFR at the onset time.

The results are given in \EFig~2. For comparison, we also plot the field profile and decay index at the initial time as well as the time when the MFR first forms. It shows that, at the onset of eruption, the apex of the newly formed MFR is located at a height of $25$~Mm, much lower than the critical height of torus instability ($\sim 75$~Mm), i.e., the height at which the decay index of the external field reaches the canonical threshold of 1.5. This suggests that the torus instability does not occur at the onset of the eruption. We note that the current in the CS below the MFR contributes a large portion to the external field, which is thus much stronger than the initial potential field (by at least several times at height above 50~Mm). Therefore, a simple approximation of the external field by the potential field, as often used in observational studies, might significantly underestimate the actual value if the current below the MFR is strong. We also note that the profile of decay index at the time when the MFR first forms (the blue line) is very close to the one (the pink line) at $t=217$~min before the MFR forms, because the current in the MFR at its initial formation is much weaker than that in the lower CS, and thus contributes very minimally to the total overlying field.

We further analyze the process of the MFR acceleration during the eruption. In \SFig~10 (and \Movie~7), we show the dynamics of a sequence of magnetic field lines that undergo reconnection and become part of the MFR. By following the movement of each field line, one can clearly see how the MFR is accelerated. After each field line approaches the CS and reconnects, its middle point (i.e., the intersection point of the field line with the central cross section) is immediately accelerated upward from nearly 0 to $500 \sim 1000$ km~s$^{-1}$, close to the local Alfv{\'e}n speed. Such acceleration is realized through the slingshot effect of reconnection by the upward magnetic tension force (\SFig~11). It is extremely rapid, reaching up to $\sim 50$ km~s$^{-2}$, and is accomplished in a few tens of seconds within a rather lower height (below 50~Mm, \SFig~10D). Then the field line is decelerated briefly because it relaxes quickly from upward to downward concave one, during which the tension force changes sign.

The high-speed magnetic flux is also decelerated as it pushes the field lines ahead in the MFR, which results in a downward magnetic pressure force for a brief interval immediately after the reconnection (\SFig~11). Specifically, the field lines that join the MFR in different time show different behaviors. The very early reconnected field line obtained relatively lower speed by the slingshot effect (for instance, the field line 1 shown in \SFig~10 obtains $400$~km~s$^{-1}$), while the later-reconnected ones gain higher and higher speed (e.g., 500~km~s$^{-1}$ for field line 2, 700~km~s$^{-1}$ for field line 3, and so on) with the rise of the reconnection rate (and the energy conversion rate) until its peak time ($t = 227$~min). Therefore, the early reconnected one is pushed upward by the later reconnected, faster ones from below, and thus is accelerated, and this process occurs one-by-one, seamlessly, for each field line that joins the MFR in time sequence. This secondary phase of acceleration of the field lines, with values of $\sim 1$~km~s$^{-2}$, is much slower than the initial one by the slingshot effect of reconnection, and it continues much longer until the reconnection rate reaches its peak (see especially the field line 1). Eventually, all the field lines approach a nearly uniform speed of $500\sim 600$~km~s$^{-1}$ at which the MFR erupts as a whole (\SFig~10).

Thus, for each field line of the MFR, the initial acceleration by reconnection plays the key role in determining its final speed, and moreover, the speed achieved directly from reconnection is much higher than the final erupting speed of the MFR (except the very early reconnected ones). This analysis clearly suggests that the reconnection with its slingshot effect is the central engine of MFR acceleration, or in other words, most of the work for the MFR acceleration is done by the upward tension force as a result of the reconnection. The bursty nature of such acceleration and its accomplishment within a rather lower height is inherent to the reconnection of the strong field. This is unlike MFR acceleration through torus instability, which should last for a larger height and a longer time after the MFR runs across the threshold height, and thus work less impulsively than the reconnection.

However, whether the MFR, after being accelerated by the reconnection, can escape into the solar wind as a CME, should also depend on the strength of the overlying field relative to the core field as well as its decay rate with height. Our simulation shows that in the later phase, i.e., after the peak time of energy conversion rate, the hoop force of the MFR (approximately the magnetic pressure force) approximately balances the strapping force (approximately the tension force, see \SFig~11), indicating that the MFR is close to torus instability. Such behavior is relevant to details of the overlying or envelope field of the sheared core. We anticipate that a stronger overlying field configuration might render the eruption failed, if the strapping force exceeds the hoop force of the MFR. With a further stronger overlying field, the reconnection might be slowed and terminated very early by the strong confinement, simply producing a confined flare without formation of MFR.
%which is also proposed in the original tether-cutting scenario~\citep{Moore2001}.
On the contrary, a weaker (and faster decaying) overlying field could allow the MFR to run into torus instability, which will lead to even faster eruption. Similarly, if the overlying field consists of a multipolar configuration with a null point, as required in the breakout model, reconnection at the null also provides an efficient way to weaken the overlying field, thus facilitating the MFR to escape.

\clearpage

%\bibliographystyle{unsrt}
%\bibliographystyle{aasjournal}
%\bibliography{all}

\begin{thebibliography}{10}

\bibitem{Fleishman2020}
Fleishman, G.~D., et al.
\newblock Decay of the coronal magnetic field can release sufficient energy to
  power a solar flare.
\newblock {\em Science}, 367, 278--280 (2020).

\bibitem{Priest2002}
{Priest}, E.~R. \& {Forbes}, T.~G.
\newblock {The magnetic nature of solar flares}.
\newblock {\em \aapr}, 10, 313--377 (2002).

\bibitem{Forbes2006}
{Forbes}, T.~G., et al.
\newblock {CME Theory and Models}.
\newblock {\em \ssr}, 123, 251--302 (2006).

\bibitem{Shibata2011}
{Shibata}, K. \& {Magara}, T.
\newblock {Solar flares: magnetohydrodynamic processes}.
\newblock {\em Living Rev. Sol. Phys.}, 8, 6 (2011).

\bibitem{ChenP2011}
{Chen}, P.~F.
\newblock {Coronal mass ejections: models and their observational basis}.
\newblock {\em Living Rev. Sol. Phys.}, 8, 1 (2011).

\bibitem{Schmieder2013}
{Schmieder}, B., {D{\'e}moulin}, P. \& {Aulanier}, G.
\newblock {Solar filament eruptions and their physical role in triggering
  coronal mass ejections}.
\newblock {\em Advances in Space Research}, 51, 1967--1980 (2013).

\bibitem{Aulanier2014}
{Aulanier}, G.
\newblock {The physical mechanisms that initiate and drive solar eruptions}.
{\em IAU Symposium}, 300, 184--196 (2014).

\bibitem{Janvier2015}
{Janvier}, M., {Aulanier}, G. \& {D{\'e}moulin}, P.
\newblock {From coronal observations to MHD simulations, the building blocks
  for 3D models of solar flares (invited review)}.
\newblock {\em \solphys}, 290, 3425--3456 (2015).

\bibitem{LinJ2015}
Lin, J., et al.
\newblock Review on current sheets in cme development: theories and
  observations.
\newblock {\em \ssr}, 194, 237--302 (2015).

\bibitem{Kliem2006}
{Kliem}, B. \& {T{\"o}r{\"o}k}, T.
\newblock {Torus instability}.
\newblock {\em Phys. Rev. Lett.}, 96, 255002 (2006).

\bibitem{Torok2005}
{T{\"o}r{\"o}k}, T. \& {Kliem}, B.
\newblock {Confined and ejective eruptions of kink-unstable flux ropes}.
\newblock {\em \apjl}, 630, L97--L100 (2005).

\bibitem{Fan2007}
{Fan}, Y. \& {Gibson}, S.~E.
\newblock {Onset of coronal mass ejections due to loss of confinement of
  coronal flux ropes}.
\newblock {\em \apj}, 668, 1232--1245 (2007).

\bibitem{Aulanier2010}
{Aulanier}, G., {T{\"o}r{\"o}k}, T., {D{\'e}moulin}, P. \& {DeLuca}, E.~E.
\newblock {Formation of torus-unstable flux ropes and electric currents in
  erupting sigmoids}.
\newblock {\em \apj}, 708, 314--333 (2010).

\bibitem{Amari2018}
{Amari}, T., {Canou}, A., {Aly}, J.-J., {Delyon}, F. \& {Alauzet}, F.
\newblock {Magnetic cage and rope as the key for solar eruptions}.
\newblock {\em \nat}, 554, 211--215 (2018).

\bibitem{Antiochos1999}
{Antiochos}, S.~K., {DeVore}, C.~R. \& {Klimchuk}, J.~A.
\newblock {A model for solar coronal mass ejections}.
\newblock {\em \apj}, 510, 485--493 (1999).

\bibitem{Aulanier2000}
{Aulanier}, G., {DeLuca}, E.~E., {Antiochos}, S.~K., {McMullen}, R.~A. \&
 {Golub},  L.
\newblock {The topology and evolution of the Bastille Day flare}.
\newblock {\em \apj}, 540, 1126--1142 (2000).

\bibitem{Lynch2008}
Lynch, B.~J., Antiochos, S.~K., DeVore, C.~R., Luhmann, J.~G. \& Zurbuchen, T.~H.
\newblock Topological evolution of a fast magnetic breakout CME in
  three dimensions.
\newblock {\em \apj}, 683, 1192--1206 (2008).

\bibitem{Wyper2017}
{Wyper}, P.~F., {Antiochos}, S.~K. \& {DeVore}, C.~R.
\newblock {A universal model for solar eruptions}.
\newblock {\em \nat}, 544, 452--455 (2017).

\bibitem{Patsourakos2020}
{Patsourakos}, S., et al.
\newblock {Decoding the pre-eruptive magnetic field configurations of coronal mass ejections}.
\newblock {\em \ssr}, 216, 131 (2020).

\bibitem{DeVore2000}
{DeVore}, C.~R. \& {Antiochos}, S.~K.
\newblock {Dynamical formation and stability of helical prominence magnetic fields}.
\newblock {\em \apj}, 539, 954--963 (2000).

\bibitem{WangH2015}
Wang, H., et al. Witnessing magnetic twist with high-resolution observation from the 1.6-m New Solar Telescope. {\em Nat. Commun.}, 6, 7008 (2015).

\bibitem{WangW2017}
Wang, W., et al. Buildup of a highly twisted magnetic flux rope during a solar eruption. {\em Nat. Commun.}, 8, 1330 (2017).

\bibitem{Ugarteurra2007}
Ugarte-Urra, I., Warren, H. P. \& Winebarger, A. R. The Magnetic Topology of Coronal Mass Ejection Sources. {\em \apj}, 662, 1293--1301 (2007).

\bibitem{Moore1980}
{Moore}, R.~L. \& {Labonte}, B.~J.
\newblock {The filament eruption in the 3B flare of July 29, 1973 - Onset and
  magnetic field configuration}.
%\newblock In M.~{Dryer} and E.~{Tandberg-Hanssen}, editors, {\em Solar and Interplanetary Dynamics},
{\em IAU Symposium}, 91, 207--210 (1980).

\bibitem{Moore1992}
{Moore}, R.~L. \& {Roumeliotis}, G.
\newblock {Triggering of eruptive flares - destabilization of the preflare magnetic field configuration}.
%\newblock In Z.~{Svestka}, B.~V. {Jackson}, and M.~E. {Machado}, editors, {\emIAU Colloq. 133: Eruptive Solar Flares},
{\em Lecture Notes in Physics, Berlin Springer Verlag}, 399, 69 (1992).

\bibitem{Moore2001}
{Moore}, R.~L., {Sterling}, A.~C., {Hudson}, H.~S. \& {Lemen}.  J.~R.
\newblock {Onset of the magnetic explosion in solar flares and coronal mass ejections}.
\newblock {\em \apj}, 552, 833--848 (2001).

\bibitem{Schrijver2007}
Schrijver, C. J. A Characteristic Magnetic Field Pattern Associated with All Major Solar Flares and Its Use in Flare Forecasting. {\em \apj}, 655, L117--L120 (2007).

\bibitem{Toriumi2019}
Toriumi, S. \& Wang, H. Flare-productive active regions. {\em Living Rev. Sol. Phys.}, 16, 3 (2019).


\bibitem{Emslie2012}
Emslie, A.~G., et al.
\newblock {Global} {energetics} {of} {thirty}-{eight} {large} {solar} {eruptive} {events}.
\newblock {\em \apj}, 759, 71 (2012).

\bibitem{ZhangJ2001}
Zhang, J., Dere, K. P., Howard, R. A., Kundu, M. R. \& White, S. M. On the Temporal Relationship between Coronal Mass Ejections and Flares. {\em \apj}, 559, 452--462 (2001).

\bibitem{ZhangJ2006}
Zhang, J. \& Dere, K. P. A Statistical Study of Main and Residual Accelerations of Coronal Mass Ejections. {\em \apj}, 649, 1100--1109 (2006).


\bibitem{ChengX2020}
Cheng, X. et al. Initiation and Early Kinematic Evolution of Solar Eruptions. {\em \apj}, 894, 85 (2020).

\bibitem{Aly1991}
Aly, J.~J.
\newblock How much energy can be stored in a three-dimensional force-free magnetic field?
\newblock {\em \apj}, 375, L61--L64, (1991).

\bibitem{Sturrock1991}
Sturrock, P.~A.
\newblock Maximum energy of semi-infinite magnetic field configurations.
\newblock {\em \apj}, 380, 655-659, (1991).

\bibitem{Petschek1964}
{Petschek}, H.~E.
\newblock {\em {Magnetic Field Annihilation}}, 50, 425 (1964).

\bibitem{Linker2003}
Linker, J. A., et al. Flux cancellation and coronal mass ejections. {\em Phys. Plasmas}, 10:1971--1978, (2003).

\bibitem{Amari2003A}
{Amari}, T.,  {Luciani}, J.~F.,  {Aly}, J.~J., {Mikic}, Z. \& {Linker}, J.
\newblock {Coronal Mass Ejection: Initiation, Magnetic Helicity, and Flux
  Ropes. I. Boundary Motion-driven Evolution}.
\newblock {\em \apj}, 585, 1073--1086 (2003).

\bibitem{Torok2018}
{T{\"o}r{\"o}k}, T., et al.
\newblock {Sun-to-Earth MHD Simulation of the 2000 July 14 Bastille Day
  Eruption}.
\newblock {\em \apj}, 856, 75 (2018).

\bibitem{WangH2003}
Wang, H., Qiu, J., Jing, J. \& Zhang, H. Study of Ribbon Separation of a Flare Associated with a Quiescent Filament Eruption. {\em \apj}, 593, 564--570 (2003).

\bibitem{Hinterreiter2018}
Hinterreiter, J., Veronig, A. M., Thalmann, J. K., Tschernitz, J. \& {P{\"o}tzi}, W. Statistical Properties of Ribbon Evolution and Reconnection Electric Fields in Eruptive and Confined Flares. {\em \solphys}, 293, 38 (2018).

\bibitem{YanX2018}
{Yan}, X.~L., et al.
\newblock {Successive X-class Flares and Coronal Mass Ejections Driven by
  Shearing Motion and Sunspot Rotation in Active Region NOAA 12673}.
\newblock {\em \apj}, 856, 79 (2018).

\bibitem{Bhattacharjee2009}
Bhattacharjee, A., Huang, Y.~M., Yang, H. \& Rogers, B.
\newblock Fast reconnection in high-{Lundquist}-number plasmas due to the
  plasmoid {instability}.
\newblock {\em Phys. Plasmas}, 16, 112102 (2009).

\bibitem{Huang2010}
Huang, Y.~M. \& Bhattacharjee, A.
\newblock Scaling laws of resistive magnetohydrodynamic reconnection in the
  high-{Lundquist}-number, plasmoid-unstable regime.
\newblock {\em Phys. Plasmas}, 17, 062104 (2010).

\bibitem{Daughton2011}
Daughton, W.
\newblock Role of electron physics in the development of turbulent magnetic
  reconnection in collisionless plasmas.
\newblock {\em Nat. Phys.}, 7, 539--542 (2011).

\bibitem{Nishida2013}
Nishida, K., Nishizuka, N. \& Shibata, K.
\newblock {The} {role} {of} {a} {flux} {rope} {ejection} {in} {a}
  {three}-{dimensional} {magnetohydrodynamic} {simulation} {of} {a} {solar}
  {flare}.
\newblock {\em \apj}, 775, L39 (2013).

\bibitem{Mikic1994}
{Mikic}, Z. \& {Linker}, J.~A.
\newblock {Disruption of coronal magnetic field arcades}.
\newblock {\em \apj}, 430, 898--912 (1994).

\bibitem{Choe1996}
{Choe}, G.~S. \& {Lee}, L.~C.
\newblock {Evolution of Solar Magnetic Arcades. I. Ideal MHD Evolution under Footpoint Shearing}.
\newblock {\em \apj}, 472, 360--371 (1996).

\bibitem{Amari2003}
{Amari}, T., {Luciani}, J.~F., {Aly}, J.~J., {Mikic}, Z. \& {Linker}, J.
\newblock {Coronal Mass Ejection: Initiation, Magnetic Helicity, and Flux Ropes. I. Boundary Motion-driven Evolution}.
\newblock {\em \apj}, 585, 1073--1086 (2003).

\bibitem{Karpen2012}
Karpen, J.~T., Antiochos, S.~K. \& DeVore, C.~R.
\newblock {The} {mechanisms} {for} {the} {onset} {and} {explosive} {eruption}
  {of} {coronal} {mass} {ejections} {and} {eruptive} {flares}.
\newblock {\em \apj}, 760, 81 (2012).

\bibitem{Yardley2018}
Yardley, S.~L., Green,   L.~M., van Driel-Gesztelyi,  L., Williams, D.~R. \& Mackay, D.~H.
\newblock The {role} of {flux} {cancellation} in {eruptions} from {bipolar}{ARs}.
\newblock {\em \apj}, 866, 8 (2018).

\bibitem{Ballegooijen1989}
{van Ballegooijen}, A.~A. \& {Martens}, P.~C.~H.
\newblock {Formation and eruption of solar prominences}.
\newblock {\em \apj}, 343, 971--984 (1989).

\bibitem{Jiang2010}
Jiang, C.~W., Feng, X.~S., Zhang, J. \& Zhong, D.~K.
\newblock {AMR simulations of magnetohydrodynamic problems by the CESE method in curvilinear coordinates}.
\newblock {\em \solphys}, 267, 463--491 (2010).

\bibitem{Feng2010}
{Feng}, X.~S., et al.
\newblock {Three-dimensional solar wind modeling from the sun to earth by a
  sip-cese mhd model with a six-component grid}.
\newblock {\em \apj}, 723, 300--319 (2010).

\bibitem{Jiang2016NC}
{Jiang}, C.~W., {Wu}, S.~T., {Feng}, X.~S. \& {Hu}, Q.
\newblock Data-driven MHD simulation of a flux-emerging active region leading to solar eruption.
\newblock {\em Nat. Commun.}, 7, 11522 (2016).

\bibitem{Brown2003}
{Brown}, D.~S., et al.
\newblock {Observations of Rotating Sunspots from TRACE}.
\newblock {\em \solphys}, 216, 79--108 (2003).

\bibitem{YanX2007}
{Yan}, X.~L. \& {Qu},  Z.~Q.
\newblock {Rapid rotation of a sunspot associated with flares}.
\newblock {\em \aap}, 468, 1083--1088 (2007).

\bibitem{YanX2012}
{Yan}, X.~L., {Qu}, Z.~Q., {Kong}, D.~F. \& {Xu}, C.~L.
\newblock {Sunspot Rotation, Sigmoidal Filament, Flare, and Coronal Mass Ejection: The Event on 2000 February 10}.
\newblock {\em \apj}, 754, 16 (2012).

\bibitem{Amari1996B}
{Amari}, T., {Luciani}, J.~F., {Aly}, J.~J. \& {Tagger}, M.
\newblock {Very Fast Opening of a Three-dimensional Twisted Magnetic Flux Tube}.
\newblock {\em \apjl}, 466, L39--L42 (1996).

\bibitem{Tokman2002}
{Tokman}, M. \& {Bellan}, P.~M.
\newblock {Three-dimensional Model of the Structure and Evolution of Coronal Mass Ejections}.
\newblock {\em \apj}, 567, 1202--1210 (2002).

\bibitem{Torok2003}
{T{\"o}r{\"o}k}, T. \& {Kliem}, B.
\newblock {The evolution of twisting coronal magnetic flux tubes}.
\newblock {\em \aap}, 406, 1043--1059 (2003).

\bibitem{DeVore2008}
{DeVore}, C.~R. \& {Antiochos}, S.~K.
\newblock {Homologous Confined Filament Eruptions via Magnetic Breakout}.
\newblock {\em \apj}, 680, 740--756 (2008).

\bibitem{Shibata2001}
Shibata, K. \& Tanuma, S.
\newblock Plasmoid-induced-reconnection and fractal reconnection.
\newblock {\em Earth, Planets and Space}, 53, 473--482 (2001).

\bibitem{Priest1987Book}
{Priest}, E.~R.
\newblock {\em {Solar magneto-hydrodynamics.}}
\newblock Springer Netherlands (1987).

\bibitem{Shiota2008}
Shiota, D., Kusano, K., Miyoshi, T., Nishikawa, N. \& Shibata, K.
\newblock A quantitative {MHD} study of the relation among arcade shearing,
  flux rope formation, and eruption due to the tearing instability.
\newblock {\em \jgr: Space Physics}, 113, A03S05 (2008).

\bibitem{Jiang2016ApJ}
Jiang, C., et al.
\newblock How did a major confined flare occur in super solar active region 12192?
\newblock {\em \apj}, 828, 62 (2016).

\bibitem{Spitzer1962}
Spitzer, L.
\newblock {\em Physics of fully ionized gas (2nd edition)}.
\newblock Interscience, New York (1962).

\bibitem{Yokoyama1994}
Yokoyama, T. \& Shibata, K.
\newblock What is the condition for fast magnetic reconnection?
\newblock {\em \apj}, 436, L197--L200 (1994).

\bibitem{Lazarian1999}
Lazarian, A. \& Vishniac, E.~T.
\newblock Reconnection in a {weakly} {stochastic} {field}.
\newblock {\em \apj}, 517, 700--718 (1999).

\bibitem{Kowal2009}
Kowal, G., Lazarian, A., Vishniac, E.~T. \& Otmianowska-Mazur, K.
\newblock {Numerical} {tests} {of} {fast} {reconnection} {in} {weakly} {stochastic} {magnetic} {fields}.
\newblock {\em \apj}, 700, 63--85 (2009).

\bibitem{Aulanier2012}
{Aulanier}, G., {Janvier}, M. \& {Schmieder}, B.
\newblock {The standard flare model in three dimensions. I. Strong-to-weak shear transition in post-flare loops}.
\newblock {\em \aap}, 543, A110 (2012).

\bibitem{Janvier2013}
Janvier, M., Aulanier, G., Pariat, E. \& D{\'e}moulin, P.
\newblock The standard flare model in three dimensions: {III}. {Slip}-running reconnection properties.
\newblock {\em \aap}, 555, A77 (2013).

\bibitem{Inoue2013}
{Inoue}, S., {Hayashi}, K., {Shiota}, D., {Magara}, T. \& {Choe}, G.~S.
\newblock {Magnetic Structure Producing X- and M-class Solar Flares in Solar Active Region 11158}.
\newblock {\em \apj}, 770, 79 (2013).

\bibitem{Savcheva2016}
{Savcheva}, A., et al.
\newblock {The Relation between Solar Eruption Topologies and Observed Flare
  Features. II. Dynamical Evolution}.
\newblock {\em \apj}, 817, 43 (2016).

\bibitem{LiuR2016}
Liu, R., et al.
\newblock Structure, stability, and evolution of magnetic flux ropes from the
  perspective of magnetic twist.
\newblock {\em \apj}, 818, 148 (2016).

\bibitem{DuanA2019}
Duan, A., et al. A Study of Pre-flare Solar Coronal Magnetic Fields: Magnetic Flux Ropes. {\em \apj}, 884, 73 (2019).

\bibitem{Titov2002}
{Titov}, V.~S., {Hornig}, G., \& {D{\'e}moulin}, P.
\newblock {Theory of magnetic connectivity in the solar corona}.
\newblock {\em \jgr}, 107, 1164, 2002.

\bibitem{Berger2006}
{Berger}, M.~A. \& {Prior}, C.
\newblock {The writhe of open and closed curves}.
\newblock {\em Journal of Physics A Mathematical General}, 39, 8321--8348 (2006).

\bibitem{QiuJ2002}
{Qiu}, J., {Lee}, J., {Gary}, D.~E. \& {Wang}, H.~M.
\newblock {Motion of Flare Footpoint Emission and Inferred Electric Field in Reconnecting Current Sheets}.
\newblock {\em \apj}, 565, 1335--1347 (2002).

\bibitem{JiangC2018}
Jiang, C., et al. Magnetohydrodynamic simulation of the X9.3 flare on 2017 September 6: Evolving magnetic topology. {\em \apj}, 869, 13 (2018).

\bibitem{QiuJ2017}
Qiu, J., Longcope, D. W., Cassak, P. A. \& Priest, E. R. Elongation of Flare Ribbons. {\em \apj}, 838, 17 (2017).

\bibitem{SuY2007}
Su, Y., Golub, L. \& Van Ballegooijen, A. A. A statistical study of shear motion of the footpoints in two-ribbon flares. {\em \apj}, 655, 606--614 (2007).

\bibitem{Demoulin1996JGR}
{D{\'e}moulin}, P., {Priest}, E.~R. \& {Lonie}, D.~P.
\newblock {Three-dimensional magnetic reconnection without null points: 2. Application to twisted flux tubes}.
\newblock {\em \jgr}, 101, 7631--7646 (1996).

\bibitem{Savcheva2012a}
{Savcheva}, A., {Pariat}, E., {van Ballegooijen}, A., {Aulanier}, G. \& {DeLuca}. E.
\newblock {Sigmoidal Active Region on the Sun: Comparison of a Magnetohydrodynamical Simulation and a Nonlinear Force-free Field Model}.
\newblock {\em \apj}, 750, 15 (2012).

\bibitem{Janvier2014}
Janvier, M.
\newblock Electric currents in flare ribbons: Observations and
  three-dimensional standard model.
\newblock {\em \apj}, 788, 60 (2014).

\bibitem{JingJ2016}
Jing, J., et al. Unprecedented Fine Structure of a Solar Flare Revealed by the 1.6 m New Solar Telescope. {\em Scientific Reports}, 6, 24319 (2016).

\end{thebibliography}
%\bibliographystyle{apj}%aasjournal}

\clearpage

\renewcommand\thefigure{\Alph{section}\arabic{figure}}
\setcounter{figure}{0}

\begin{figure*}[htbp]
   \centering
   \includegraphics[width=\textwidth]{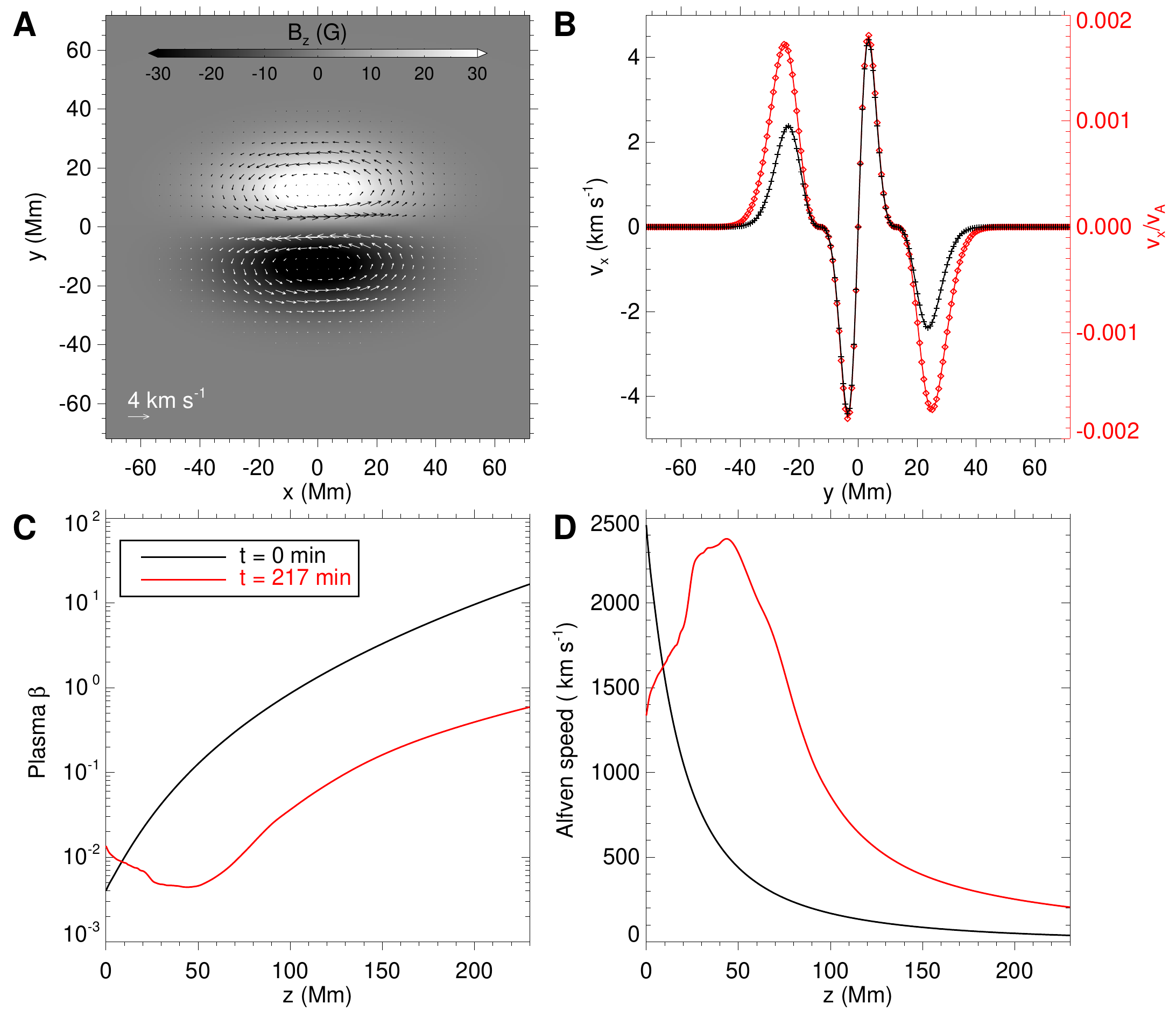}
   \caption*{\textbf{\EFig 1. Some key parameters for the settings of the simulation}.
   (\textbf{A}) Magnetic flux distribution and surface
     rotation flow at the bottom surface (i.e., $z=0$). The background
     is color-coded by the vertical magnetic component $B_{z}$ , and
     the vectors show the rotation flow.  (\textbf{B}) Profile of
     velocity (the black line) and its ratio to local Alfv{\'e}n speed (the red line) along $(x, z)=0$
     line.  (\textbf{C}) Plasma $\beta$ (i.e., ratio of gas pressure
     to the magnetic pressure) profile along the central vertical
     line, i.e., $(x,y)=0$. (\textbf{D}) Profile of Alfv{\'e}n speed
     along the central vertical line. In (C) and (D), the black lines
     are shown for the initial values, while the red lines represent
     the values at time immediately prior to the eruption onset. }
   \label{S1}
\end{figure*}

\begin{figure*}[htbp]
  \centering
  \includegraphics[width=\textwidth]{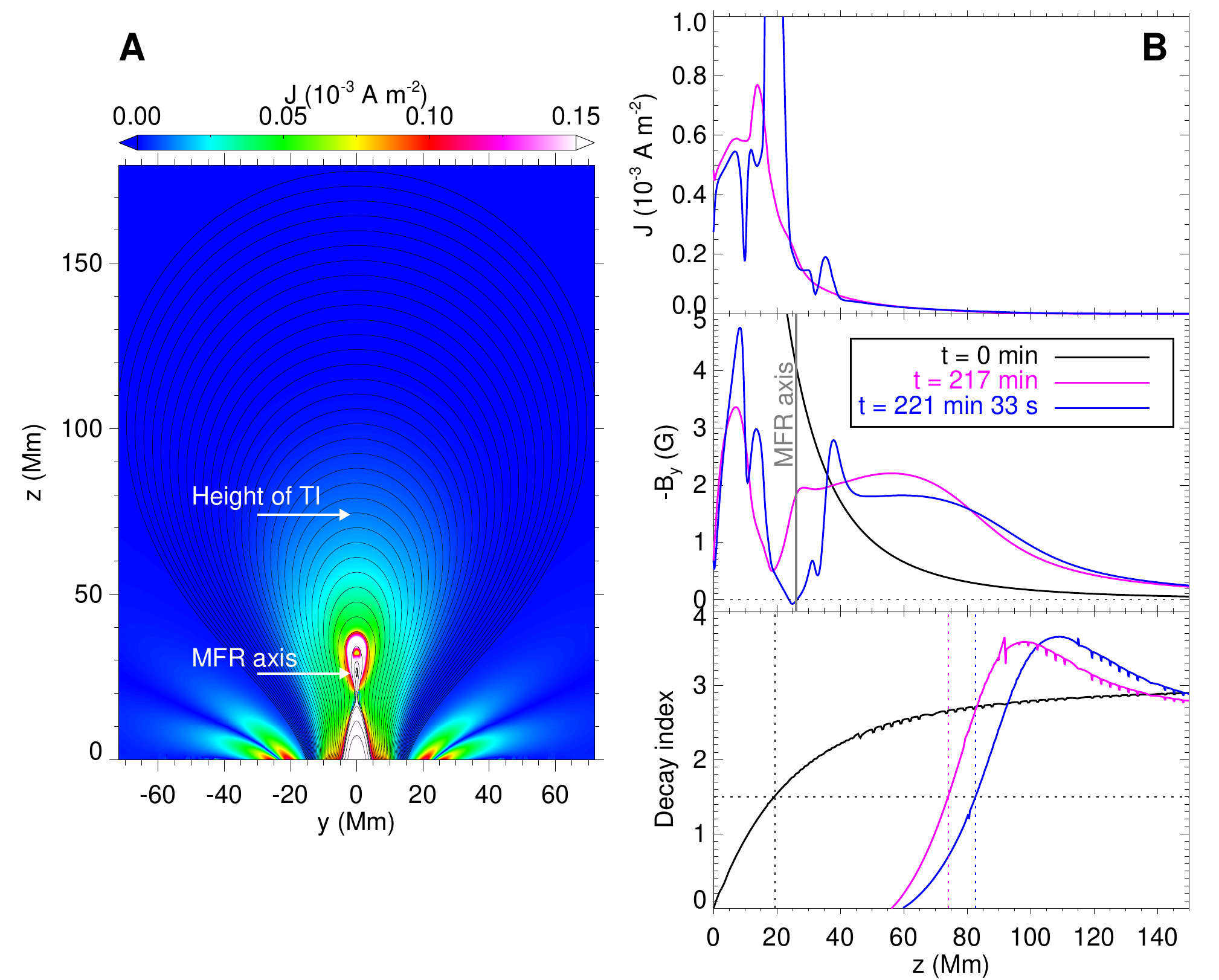}
  \caption*{\textbf{\EFig 2. Magnetic field, current density and decay index around the eruption onset}.
  (\textbf{A}) Current density on the slice of $x=0$ at the time when the MFR first
    forms during the eruption (i.e., $t=221$~min~33~s). The black curves are
    projection of magnetic field lines on the slice. The lower arrow denotes the axis of the MFR.
    The upper arrow denotes the critical height of torus instability (TI).
    %Above the MFR, the field is close to current-free and plays the role of strapping field of the MFR
    (\textbf{B}) From top to bottom are shown for current density, magnetic field component $B_y$, and decay index of $B_y$, respectively, along $z$ axis (i.e., the line with both $x$ and $y=0$). The black, magenta, and blue curves represent results for the initial potential field ($t=0$), the field immediately prior to the eruption onset ($t=217$~min), and the field at $t=221$~min~33~s, respectively. In the middle panel, the thick vertical line colored in gray denotes the height at which
    the MFR is initially formed. In the bottom panel, the dashed horizontal line denote
    the critical value ($1.5$) of decay index, and the dashed vertical lines denote the corresponding heights.
    %Clearly at the onset of the eruption, the strapping field is significantly strengthened comparing with the initial potential field, and the MFR forms well below the critical height of TI.
    }
  \label{S2}
\end{figure*}

\begin{figure*}[htbp]
  \centering
  \includegraphics[width=1\textwidth]{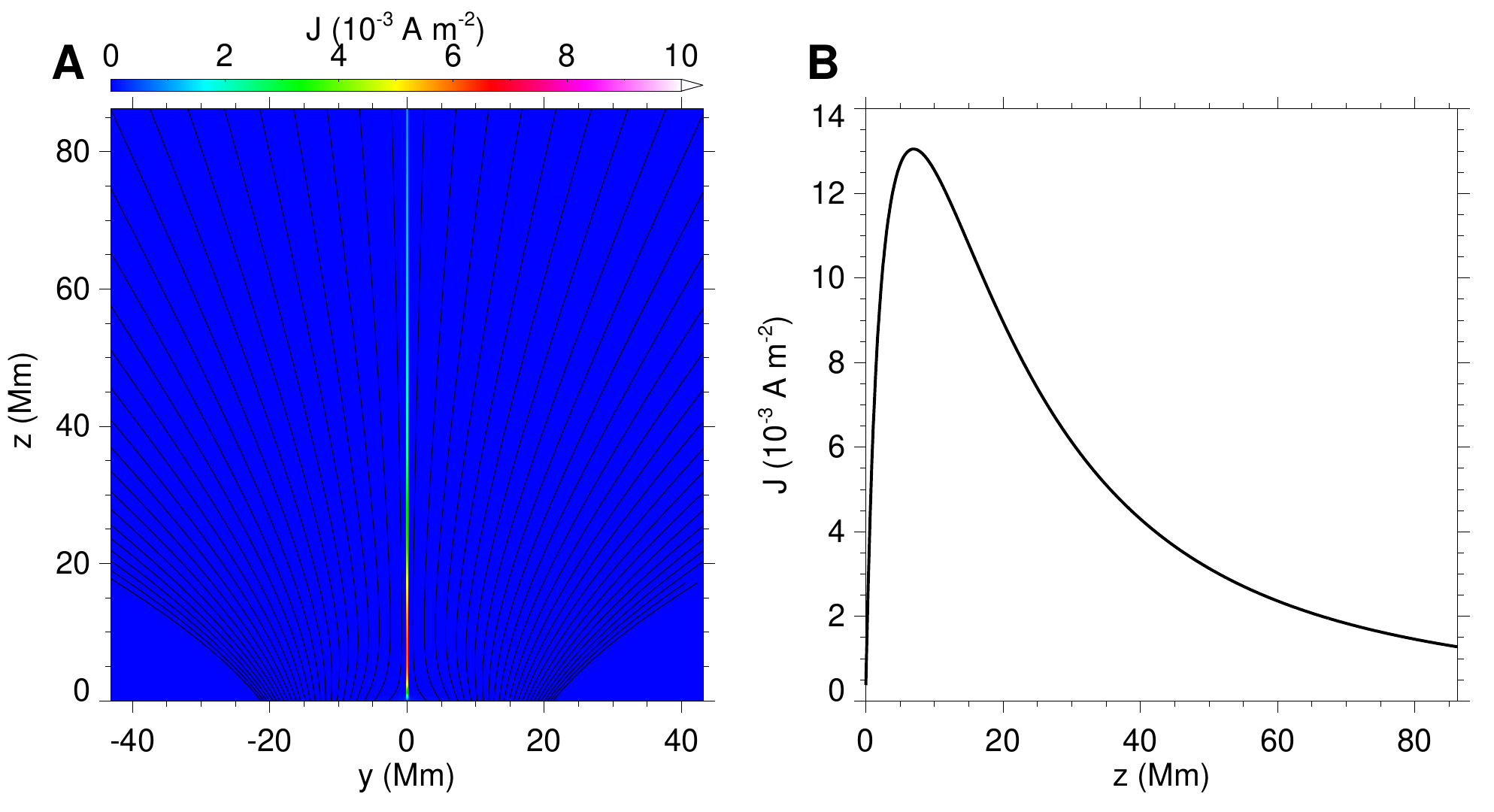}
  \caption*{\textbf{\EFig 3. The fully opened magnetic field discretized on grid with
    resolution of $90$~km}.  \textbf{(A)} Current density distribution
    on the central cross section, i.e., the $x=0$ slice, showing that current only distributes
    in the central line, or more exactly a CS with a finite thickness of $90$~km, while all other regions are current-free.
    The black curves represent the magnetic field lines, which are fully opened, i.e., extending
     from the bottom surface to infinity.  \textbf{(B)} Profile of current
    density along $z$ axis. }
  \label{S3}
\end{figure*}

\begin{figure*}[htbp]
  \centering
  \includegraphics[width=1\textwidth]{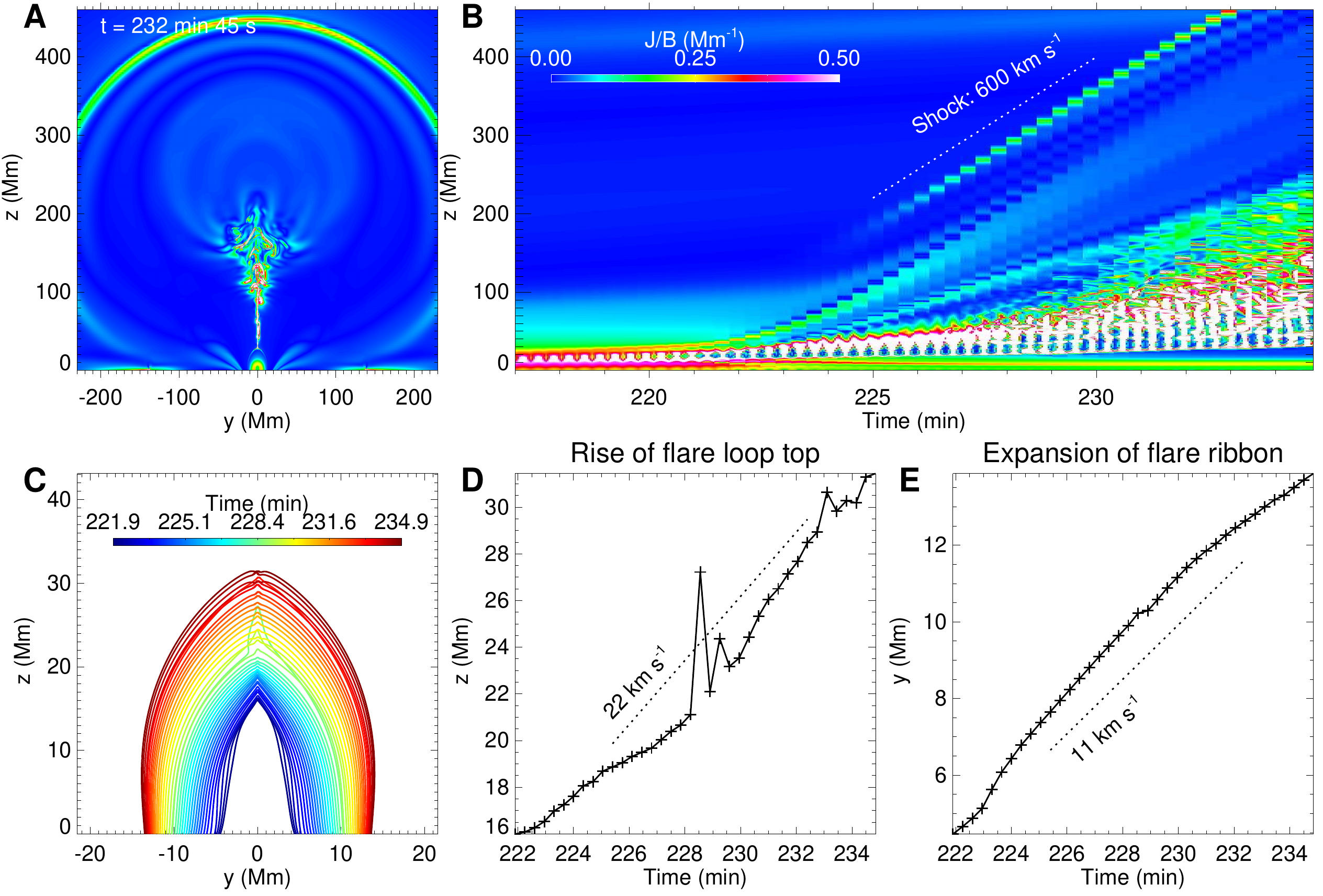}
  \caption*{\textbf{\EFig 4. Parameters that are comparable with
    observations}. \textbf{(A)} Current distribution on the central
    cross section. \textbf{(B)} A time stack map of the current
    distribution around $x,y=0$, which can reveal the evolution speed
    of the CME. \textbf{(C)} Temporal evolution of the edge of the
    post-flare loops. \textbf{(D)} Rising of the post-flare loop
    top. \textbf{(E)} Horizontal motion of the post-flare loop
    footpoints, which corresponds to the separation of flare
    ribbons. The dashed lines in \textbf{D} and \textbf{E} denote the
    average speeds of the motions.}
  \label{S4}
\end{figure*}

\begin{figure*}[htbp]
  \centering
  \includegraphics[width=0.8\textwidth]{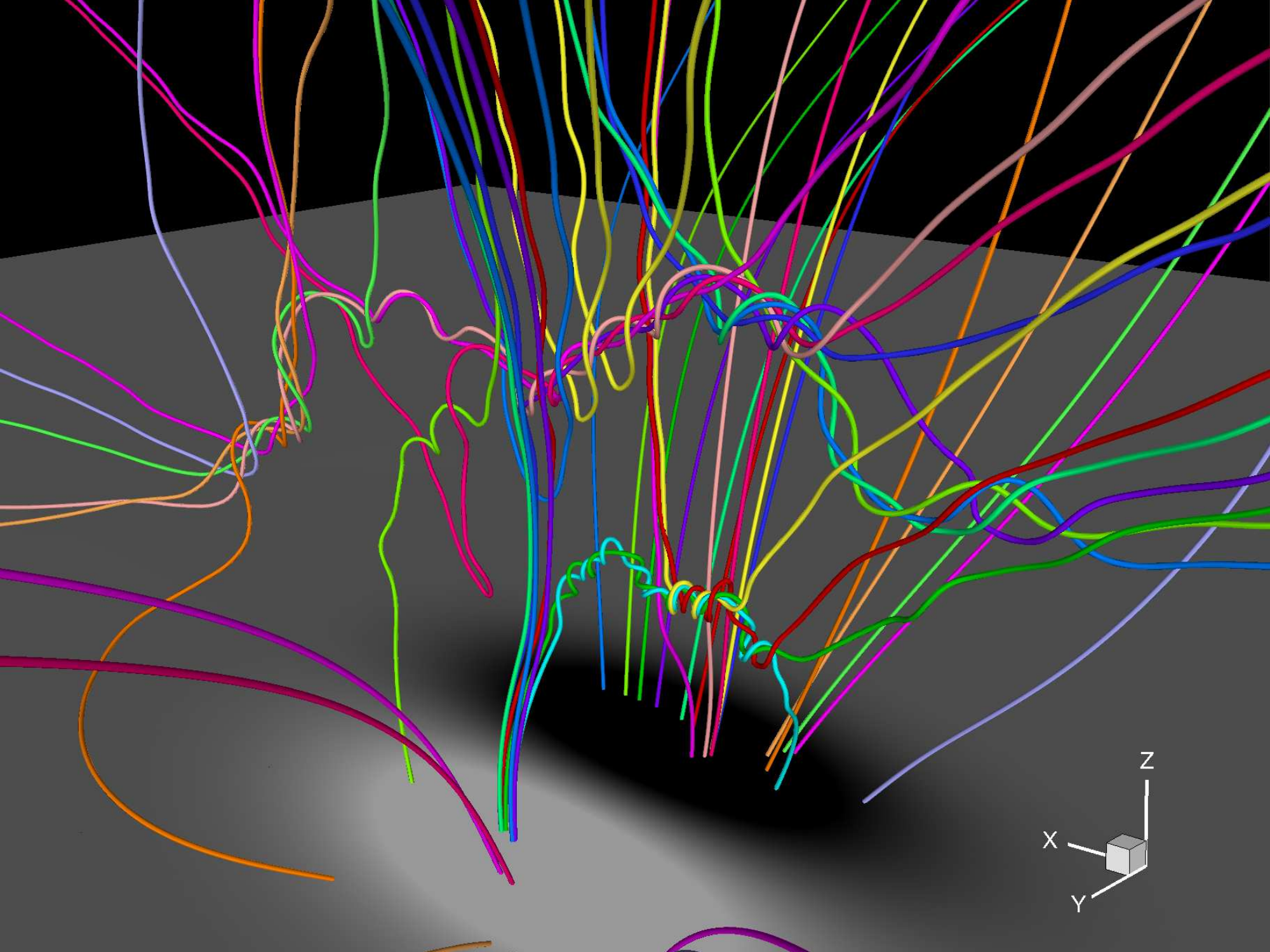}
  \caption*{\textbf{\EFig 5. Mini flux ropes formed in the reconnecting CS}. The field
    lines are colored differently and the bottom surface is shown with
    the magnetic flux distribution.}
  \label{S5}
\end{figure*}

\end{document}